\input epsf

\documentclass[12pt]{article}

\usepackage{paper2e}
\usepackage{mydefs2e}

\renewcommand{\Box}{\,\raisebox{-.45pt}{\drawsquare{7}{0.6}}\,}
\newcommand{\MP}{M_{\rm Pl}}
\newcommand{\LUV}{\La_{\rm UV}}
\renewcommand{\d}{\partial}

\begin{document}
\begin{titlepage}

\preprint{HUTP-03/A081\\UMD-PP-04-012}

\title{Ghost Condensation and a Consistent\\\medskip
Infrared Modification of Gravity}

\author{Nima Arkani-Hamed$^{\rm a}$, Hsin-Chia Cheng$^{\rm a}$,
Markus A. Luty$^{\rm a,b,c}$, Shinji Mukohyama$^{\rm a}$}

\address{$^{\rm a}$Jefferson Laboratory of Physics, Harvard University\\
Cambridge, Massachusetts 02138}

\address{$^{\rm b}$Physics Department, Boston University\\
Boston, Massachusetts 02215}

\address{$^{\rm c}$Physics Department, University of Maryland\\
College Park, Maryland 20742}

\begin{abstract}
We propose a theoretically consistent modification of gravity in
the infrared, which is  compatible with all current experimental
observations. This is an analog of the Higgs mechanism in general 
relativity, and can be thought of as arising from ghost
condensation---a background where a scalar field $\phi$ has a
constant velocity, $\langle \dot\phi \rangle = M^2$. The ghost
condensate is a new kind of fluid that can fill the universe,
which has the same equation of state, $\rho = -p$, as a
cosmological constant, and can hence drive de Sitter expansion of
the universe. However, unlike a cosmological constant, it is a
physical fluid with a physical scalar excitation, which can be
described by a systematic effective field theory at low energies.
The excitation has an unusual low-energy dispersion relation
$\omega^2 \sim \vec{k}^4 / M^2$. If coupled to matter directly, it gives
rise to small Lorentz-violating effects and a new long-range
$1/r^2$ spin dependent force. In the ghost condensate, the energy
that gravitates is not the same as the particle physics energy,
leading to the possibility of both sources that can gravitate and
{\it anti}-gravitate. The Newtonian potential is modified with an
oscillatory behavior starting at the distance scale $\MP/M^2$
and the time scale $\MP^2/M^3$. This theory opens
up a number of new avenues for attacking cosmological problems,
including inflation, dark matter and dark energy.
\end{abstract}

\end{titlepage}

\section{Introduction and Summary}
Gravity remains the most enigmatic of all the fundamental
interactions in nature. At the Planck scale, large quantum
fluctuations signal a breakdown of the effective field theory
description of gravity in terms of general relativity, leading to
the expectation that gravity and our notions of space and time
must be radically altered at short distances. Meanwhile,
continuing experimental probes of gravity at {\it large} distances
reveal strange phenomena at many length scales, from the
flattening of galactic rotation curves to the accelerating
universe. Traditionally, these phenomena are explained by invoking
new sources of matter and energy, such as dark matter and a tiny
cosmological constant. It is however worth contemplating the
possibility that gravity itself is changing in the infrared in
some way that might address these mysteries. This approach is
further motivated by the cosmological constant problem, which
seems to be associated with extreme infrared physics.

As a first step in this program, it is an interesting theoretical
challenge to find a theoretically consistent and experimentally
viable modification of gravity in the infrared (IR). Such attempts have
been made in the past~\cite{Fierz:1939ix,Brans:sx,Clayton:1998hv,Drummond:1999ut,Dvali:2000hr,
Jacobson:2001yj,Freese:2002sq,Soussa:2003vv,Wetterich:2003qb,
Carroll:2003wy,Khoury:2003aq}.
Many of these theories are in one
way or another varieties of scalar-tensor theories of gravity, and
therefore are not true modifications of gravity. For example, the
the long-range interactions between two masses in these theories is at most changed
by a constant factor. Much more interesting ideas are those that
modify gravity interactions in a dramatic way at large distances,
such as the  Fierz-Pauli theory of massive gravity~\cite{Fierz:1939ix},
and the Dvali-Gabadadze-Porrati (DGP) brane-world models with infinite
volume extra dimensions~\cite{Dvali:2000hr}. However, it is not yet entirely
clear that these theories are experimentally viable. In both of
these theories, there is an additional scalar degree of freedom,
$\pi$, interacting with matter with gravitational strength.
In the case of massive gravity it is the
helicity zero longitudinal mode of the graviton
\cite{vanDam:1970vg,Boulware:zf,Arkani-Hamed:2002sp},
and in the DGP model it is the ``brane-bending" mode \cite{Luty:2003vm}
(see also \cite{ddgv}).
Because of
the presence of the $\pi$'s, the theory reduces to scalar-tensor
theory of gravity, rather than general relativity, at distances
smaller than the length scale of IR modification.
Furthermore, the $\pi$'s become strongly coupled in the
ultraviolet (UV) at a scale $\La_{\rm UV}$ determined in terms of the
IR modification scale $\La_{\rm IR}$ and the Planck scale $\MP$.
At $\La_{\rm UV}$, the effective theory breaks down and the
physics is sensitive to the unknown UV completion of the theory.
That such a UV scale should exist is not surprising. After all,
even in non-Abelian gauge theories, if we modify the theory in the
IR by giving the gauge boson a mass $m$, this introduces a
new degree of freedom, the longitudinal component of the gauge
field. This degree of freedom becomes strongly coupled in the UV
at scale of order $4 \pi m/g$, determined by the scale of IR
modification and the gauge coupling $g$. However, the situation is
both qualitatively and quantitatively different in the
gravitational case. The highest strong coupling scale
$\La_{\rm UV}$ can be pushed in all of the above cases
is~\cite{Arkani-Hamed:2002sp,Luty:2003vm},
\beq[lowcut]
\La_{\rm UV} \sim (\La_{\rm IR}^2
\MP)^{1/3} .
\eeq
For $\La_{\rm IR} \sim H_0$, today's Hubble scale,
we have $\La_{\rm UV}^{-1} \sim 1000$ km, which is much larger than
the distances to which we have probed gravity. Roughly speaking,
this happens because gravity is being coupled to a sector that is
sick in the limit as gravity is turned off ($\MP \to \infty$).
The $\pi$'s have no kinetic term in this limit, even though
they have cubic and higher order self-interactions. The theory
only becomes tenuously healthy due to the coupling with gravity:
mixing with gravity generates small kinetic terms for $\pi$ of the
correct sign (in flat space),
but leads to strong coupling physics at low energies.
This does not necessarily mean that massive gravity or the DGP
model cannot describe the real world, only that it appears necessary
to make nontrivial assumptions about the UV
completion of the theory.

In this paper, we present a fully controllable and calculable
theory where gravity is modified in the infrared, in a way that
avoids these strong coupling issues. It is useful, though not
necessary, to view this as arising from ``ghost condensation." A
ghost condensate is a new kind of fluid that can fill the
universe.
It has the equation of state $p = -\rho$, just like a cosmological constant
$\Lambda$, and can therefore gives rise to a de Sitter phase of the
universe even if $\Lambda = 0$. But the ghost condensate is {\it
not} a cosmological constant, it is a physical fluid with a
physical scalar excitation. This background can be thought of as
arising from a theory where a real scalar field $\phi$ is changing
with a constant velocity
\begin{equation}
\langle \dot{\phi} \rangle = M^2
\quad\hbox{\rm or}\quad
\langle \phi \rangle = M^2 t ,
\end{equation}
and there is a physical scalar excitation $\pi$ around this
background,
\begin{equation}
\phi = M^2 t + \pi.
\end{equation}
We assume that the field $\phi$ has a shift
symmetry
\begin{equation}
\phi \mapsto \phi + a,
\end{equation}
so it is derivatively coupled, and the quadratic
lagrangian for $\phi$ begins with a wrong-sign kinetic term
[in $(+---)$ signature]
\begin{equation}
{\cal L}_\phi = -\sfrac 12 \partial^\mu \phi \partial_\mu \phi + \cdots\, .
\end{equation}
The wrong-sign kinetic term means that the usual background with
$\avg{\phi} = 0$ is unstable. However, as is familiar from
tachyon condensation, we will see that higher order terms can
stabilize a background where $\dot{\phi} \neq 0$. Note that this
background breaks Lorentz invariance spontaneously---there is a
preferred frame where $\phi$ is spatially isotropic. This is not
fundamentally different from the way in which the cosmic microwave
background radiation (CMBR), or any
other cosmological fluid, breaks Lorentz invariance. What is novel
about the ghost condensate is that, unlike other cosmological
fluids (but like a cosmological constant), it does not dilute as
the universe expands.\footnote{Attempts to modify gravity with
a dynamical preferred frame violating Lorentz invariance
have been made, {\it e.g.}, most recently in~\Ref{Jacobson:2001yj},
but with different physical conclusions.}

We will construct a systematic derivative expansion governing the
low-energy physics of the excitation $\pi$ about the background.
As we will see, these excitations have two important
properties, that combine to give rise to a number of surprising
new phenomena in the IR.

First, the low-energy effective action for $\pi$ has the universal
form
\begin{equation}
S \sim \myint d^4 x \left[ \frac{1}{2} \dot{\pi}^2 - \frac{1}{2 M^2}
(\nabla^2 \pi)^2 + \cdots \right],
\end{equation}
so that the $\pi$'s have a low-energy dispersion relation of the
form
\begin{equation}
\omega^2 \sim \frac{k^4}{M^2}.
\end{equation}
(Throughout this paper $k$ represents the spatial 3-momentum and
$\nabla$ represents the spatial gradient.
In this introductory section, we will not be careful about ${\cal O}(1)$
coefficients in front of various terms.)
Ordinarily the dispersion relation for light excitations is of the
form $\omega^2 \propto k^2$. Here, there is no $k^2$ term at all, so
the excitations do not move with any fixed speed. Instead, the
group velocity is $v^2 \sim k^2/M^2$, so that larger lumps of $\pi$
move more slowly than smaller lumps. Note that this is not a
Lorentz-invariant dispersion relation, as is to be expected since
the background breaks Lorentz invariance.

\begin{figure}
 \centering\leavevmode\epsfysize=8cm \epsfbox{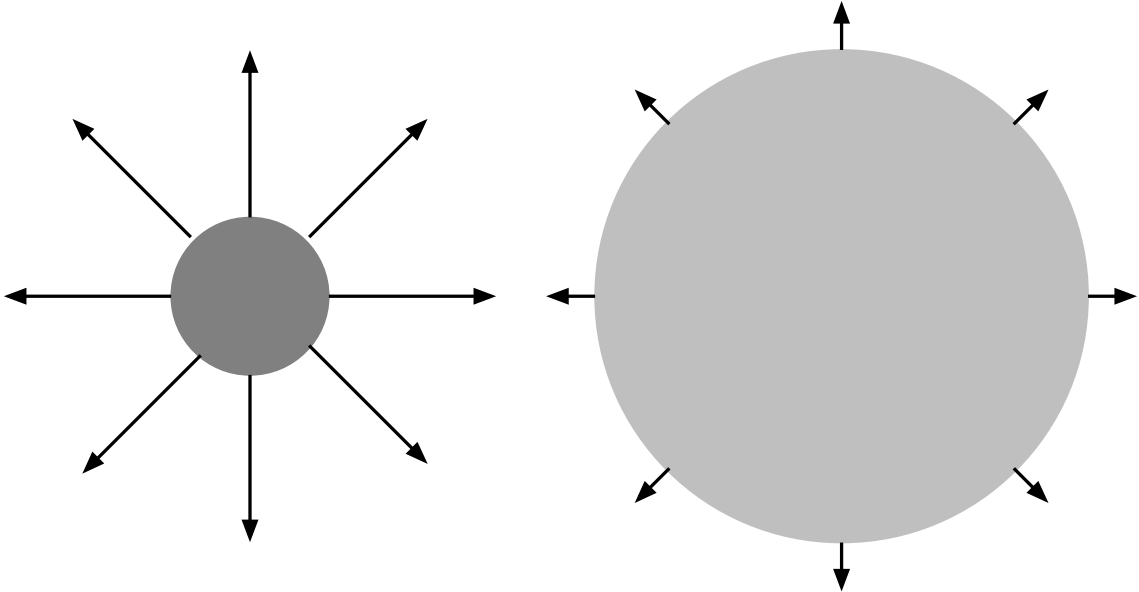}
 \caption{
 Excitations $\pi$ about the ghost condensate have a dispersion relation
 $\omega^2 \propto k^4$, so that small lumps of $\pi$ expand more
 quickly than larger lumps of $\pi$.
 }
\end{figure}

Second, there are two distinct kinds of energy in these
backgrounds: a ``particle physics" energy and a ``gravitational" energy,
and they are not the same. This happens because the background
$\langle \phi \rangle = M^2 t$ breaks time translational
invariance. It also breaks the $\phi$ shift symmetry, but the
diagonal combination is left unbroken to generate the ``time"
translations in this background. The Noether charge associated
with this unbroken symmetry is what an experimentalist coupled to
$\pi$ would call the conserved energy in this background, and this
``particle physics" energy density takes the form
\begin{equation}
{\cal E}_{\rm pp} \sim \frac{1}{2} \dot{\pi}^2 +
\frac{(\nabla^2\pi)^2}{2 M^2} + \cdots,
\end{equation}
and is manifestly positive for small fluctuations.
On the other hand, the energy density that couples to gravity
is $T_{00}$, which is associated with the broken
time-translation symmetry, and is {\it not} the same as
${\cal E}_{\rm pp}$.
Indeed,
\begin{equation}
{\cal E}_{\rm grav} = T_{00} \sim M^2 \dot{\pi} + \cdots
\end{equation}
begins at {\it linear} order in $\dot{\pi}$, therefore lumps of
$\pi$ can either gravitate or {\it anti}-gravitate depending on
the sign of $\dot{\pi}$!
In fact, the ``particle physics'' energy is nothing other than the
inertial mass, while the ``gravitational energy'' is the gravitational
mass, so we can say that the
$\pi$ excitations maximally violate the equivalence principle.
It is worth mentioning that excitations of $\pi$ with exotic gravitational
properties are not easily produced if $\pi$ couples only to gravity.
In fact, we find that turning on $\pi$ gravitationally requires long
distances and time scales, as discussed below.

\begin{figure}
 \centering\leavevmode\epsfysize=8cm \epsfbox{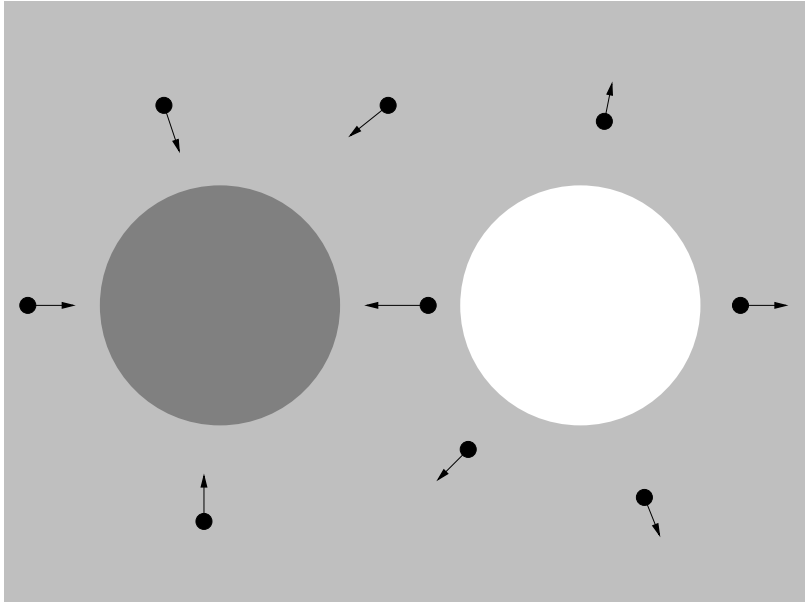}
 \caption{
 Lumps of $\pi$ can either gravitationally attract test particles (in regions
 where $\dot{\pi} > 0$) or repel them (where $\dot{\pi} < 0$).
 }
\end{figure}

The standard model fields do not need to couple directly to the
ghost sector. However, if there {\it is} a direct coupling, then
the leading interaction (a derivative coupling of $\phi$ to the
axial vector currents) gives rise to potentially observable
consequences, including a splitting between particle and
antiparticle dispersion relations, and a new spin-dependent
inverse-{\it square} law forces mediated by $\pi$ exchange. This
inverse square law behavior is a direct consequence of the
$\omega^2 \propto k^4$ dispersion relation.

Even if the standard model fields have no direct couplings to the
ghost sector, they will indirectly interact with it through
gravity, and
the propagation of gravity through the ghost condensate
gives rise to a fascinating modification of gravity in the IR. As
an example, consider a cloud of dust that quickly collapses to
form a star or a planet at some time $t=0$. At linear order, the
gravitational potential around this planet is modified at a length
scale $r_c$, but only after waiting a much longer time $t_c$, where
\begin{equation}
r_c \sim \frac{\MP}{M^2},
\qquad
t_c \sim \frac{\MP^2}{M^3}.
\end{equation}
Starting at time $t_c$ the potential becomes {\it oscillatory\/} at
distances of order $r_c$, and the modification reaches out to much
larger distances $r \gg r_c$ only after waiting an even longer
time $t \sim (r/r_c) t_c$. Furthermore, in a flat background, the
ghost condensate has a Jeans-like instability at long
wavelengths, so that the modulation of the potential begins
growing exponentially at late time $t \gg t_c$. As with ordinary
fluids, however, this Jeans instability disappears due to Hubble
friction in a cosmological background (such as de Sitter space)
with $H > t_c^{-1}$. In this case, at late time
$t \gg t_c$, the potential looks like
an oscillatory modulation of $1/r$ such as shown in Fig. 3.

\begin{figure}
 \centering\leavevmode\epsfysize=8cm \epsfbox{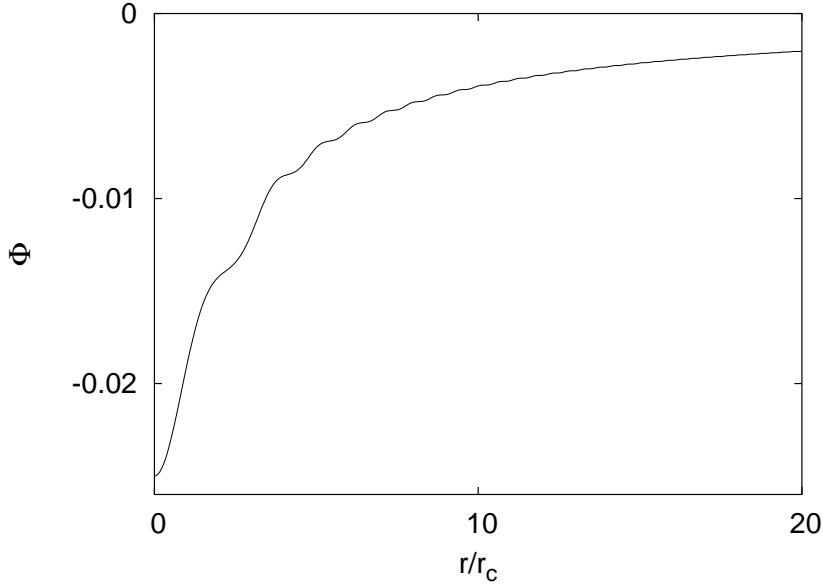}
 \caption{
 Example of a late-time Newtonian potential
 }
\end{figure}

There are a number of interesting ranges of parameters for the
mass scale $M$ associated with the ghost condensation. If the
ghost condensate is to give the observed acceleration of the
universe today, even with $\Lambda = 0$, then naturalness suggests
\begin{equation}
M \sim 10^{-3} \mbox{eV}.
\end{equation}
In this case, the distance scale where gravity is modified is $r_c
\sim H_0^{-1}$, but the time it takes to see this modification is
$t_c \gg H_{0}^{-1}$! So no modifications of gravity can be seen
directly, and no {\it cosmological\/} experiment can distinguish the
ghost-driven acceleration from a cosmological constant. If the standard model
has direct coupling to the ghost sector, we can still hope to
detect tiny Lorentz-violating effects and long-range spin-dependent
inverse square force laws. It is also possible that $M$ is much
larger; this could be the remnant of ghost condensation triggering
an earlier de Sitter epoch in the universe, to drive inflation
\cite{ghostinfl}.
For instance, we can have the parameters
\begin{equation}
M \sim 10 \, \mbox{MeV}, \quad t_c \sim H_0^{-1}, \quad r_c \sim 1000 \,
\mbox{km},
\end{equation}
so that we begin to start seeing an oscillatory modulation of
Newtonian gravity at a distance of $\sim 1000$ km, but only recently
in the history of the universe!

This physics is an exact analog of the Higgs mechanism for
gravity. Recall that general relativity is a gauging of spacetime
translations. In particular the time diffeomorphism symmetry is
spontaneously broken by our background $\langle \phi \rangle = M^2
t$, and therefore we should expect gravity to become ``massive."
There are two differences with the familiar gauge theory case.
First, our background breaks Lorentz invariance, and therefore we
do not get the massive graviton of a Lorentz invariant theory,
which would have 5 polarizations. In fact we are only adding one
additional physical degree of freedom, which can be thought of as
the scalar excitations $\pi$.
Second, while in the gauge theory case we get
a Yukawa-like exponential suppression of the potential at large
distances, in our case we get an oscillatory modulation
instead---basically because this is a {\it negative} mass-squared 
term for the
graviton. This difference in sign is to be expected in all
comparisons of gravity and gauge theory, physically arising for
the same reason that like charges repel in gauge theory and
attract in gravity. It is also analogous to the fact that fluctuations
of a charged fluid acquire a positive plasma frequency
$\omega_p^2
>0$, while gravity induces an $\omega_J^2 <0$ signalling the
Jeans instability.

We have described the physics so far in the language of gravity
propagating through a new fluid. This is like the description of
massive gauge theories in terms of propagation through the Higgs
condensate. As familiar from the gauge theory, there is also a
``unitary gauge" description of the physics, where the excitation
$\pi$ is set to zero (``eaten"), and where the interpretation of
the theory directly as an IR modification of gravity is clear. The
unitary gauge is arrived at by choosing the time coordinate $t$
such that
\begin{equation}
\phi(t,x) = M^2 t.
\end{equation}
That is, $\phi$ is taken to be the clock. This is
equivalent to making a diffeomorphism to set $\pi = 0$. In this
gauge, there are terms in the action that are not fully
diffeomorphism invariant. At quadratic level for the fluctuations
$h_{\mu \nu} = g_{\mu \nu} - \eta_{\mu \nu}$ around flat space
these new terms in the lagrangian take the form
\begin{equation}
\De{\cal L} =
 \alpha_0 M^4 h_{00}^2 - \alpha_1 M^2 K_{ij} K_{ij} -
\alpha_2 M^{2} (K_{ii})^2,
\end{equation}
where
\begin{equation}
K_{ij} = \frac12 (\partial_0 h_{ij}- \partial_i h_{0j} -\partial_j h_{0i}),
\end{equation}
and $\alpha_{0,1,2}$ are ${\cal O}(1)$ coefficients.
The fully non-linear action will be presented in section 6. These
terms are clearly ``mass" terms and unusual kinetic terms for
$h_{\mu \nu}$.

It is also instructive to understand the counting of degrees of
freedom in this theory in the $(3 + 1)$-split ADM language~\cite{ADM},
where the metric is split into $N \leftrightarrow g^{00}$, $N_i
\leftrightarrow g_{0i}$, and $\gamma_{ij} \leftrightarrow g_{ij}$.
Usually in GR, both $N$ and $N_i$ are not dynamical fields, they
are Lagrange multipliers, and their equations of motion give $1 +
3 = 4$ constraints on the $\gamma_{ij}$, reducing the $6$ degrees
of freedom in $\gamma_{ij}$ to the two physical polarizations of
the graviton. Our background breaks time-diffeomorphisms, but
leaves (time-dependent) spatial diffeomorphisms intact. As a
consequence, the $N_i$ are still Lagrange multipliers enforcing
$3$ constraints. However, $N$ is no longer a Lagrange multiplier,
and the $N$ equation of motion then simply determines $N$. There
are therefore $6 - 3 = 3$ physical degrees of freedom---we have
added one degree of freedom. Like all gauge ``symmetries,"
diffeomorphisms are not a symmetry, they are a useful redundancy
of description. The physical issue is what the degrees of freedom
and their dynamics are. The additional degree of freedom is not
manifest in the unitary gauge description of the physics, and it
is most convenient to explicitly introduce the Goldstone field $\pi$ that
non-linearly realizes the symmetry. In this way, arriving at this
theory through the background $\langle \phi \rangle= M^2 t$ of a
scalar is the analog of a linear sigma model description of gauge
symmetry breaking. This is not necessary however---we could have simply
started with the theory of the $\pi$ and its (non-linear)
transformations under time diffeomorphisms explicitly, just as we
can directly write the chiral lagrangian for pions without any
reference to an underlying linear sigma model. This line of
thought suggests another way in which our model might arise from a
more fundamental underlying theory. One of the well-known
conceptual difficulties in putting quantum mechanics and gravity
is the problem of time: quantum mechanics picks out a direction of
time, in conflict with general covariance. Obviously this problem
must be resolved at some zeroth order, however, if any remnant of
this clash remains at low energies, it will be described by our
theory at long distances and times.

In the rest of the paper we will discuss this physics in detail.
In section two we study ghost condensation in general, first in
the absence of gravity, and then in a cosmological
Friedmann-Robertson-Walker (FRW) background.
As we will see Hubble friction in an expanding universe is crucial
in determining the background. We discuss how de Sitter phases of
the universe can arise from ghost condensation in section 3.
In section 4, we discuss the low-energy effective action for
the fluctuations around the ghost condensate in the absence of
gravity, and perform the correct power-counting analysis to show
that the theory is (non-trivially) a sensible,
quantum-mechanically stable effective theory at low energies, so
the $\pi$'s are a healthy sector even in the
absence of gravity. In section 5, we discuss the signals that
would arise from a direct coupling of the standard model fields to
the ghost sector. In section 6, we systematically construct the
low-energy effective theory for the ghost condensate coupled to
gravity. In section 7, we study the modification of gravity in
the ghost condensate in flat space at linear order, arising from
mixing between the scalar sector of gravity and the $\pi$
excitations. Here we uncover the Jeans instability of the fluid at
large wavelengths, and the oscillatory modulation of the Newtonian
potential at large distances and late times. We also make some
brief comments on the interesting question of corrections to
non-linear gravity. In section 8, we repeat this analysis for
a de Sitter background, where the Jeans instability is removed. We
end with a discussion of some of the many open questions and avenues for
further research.

\section{Ghost Condensation}
\subsection{Generalities}
Consider a scalar field $\phi$ with a shift
symmetry $\phi \mapsto \phi + a$ ensuring that it always appears
derivatively in the action.
Assume that it has a wrong-sign quadratic kinetic term, {\it i.e.},
it is a {\it ghost} field:
\begin{equation}
{\cal L} = -\sfrac 12 \d^\mu \phi \d_\mu \phi + \cdots.
\end{equation}
Actually, the overall sign of the lagrangian is completely
unphysical, so what is actually relevant is that once a choice has
been made for the sign of the lagrangian for, say, the standard model and
gravity sector, a ghost is a field with opposite sign kinetic
term. The difficulty is then that, with both kinds of fields, the
energy of the theory is unbounded from below, and therefore the
vacuum $\avg\phi = 0$ is unstable. Vacuum decay will inevitably happen
quantum-mechanically. Consider the minimal case where $\phi$ is
only coupled gravitationally. Kinematically, nothing prevents the
vacuum from decaying to a pair of $\phi$'s and a pair of
gravitons. There is more and more phase space for this for higher
energy gravitons (and more negative energy $\phi$'s), but if
the $\phi$ theory has a UV cutoff at a scale $M$, then the rate
for vacuum decay per unit volume is
$\Gamma \sim {\LUV^8}/{\MP^4}$.
For $\LUV \gsim 10^{-3} \eV$, this decay
will happen within  our Hubble volume within a Hubble time.
Other possible decays and constraints were considered
in~\Ref{Carroll:2003st,Cline:2003gs}.

Let us compare this with the more familiar example of a tachyon,
which is a scalar with negative mass squared.
In this case the vacuum $\avg{\phi} = 0$ is also unstable.
Even if classically we can sit at the point $\phi=0$ forever,
quantum-mechanically we inevitably roll off the top of the
potential with a time scale of order $m^{-1}$.
In the tachyon case, we are accustomed to the idea of tachyon
condensation:
if there are higher order terms that turn the potential around, \eg,
$V(\phi) = -\sfrac 12 m^2 \phi^2 + \sfrac 14 \lambda \phi^4 + \cdots$
with $\la > 0$,
then there can be a stable vacuum where $\langle \phi \rangle \neq 0$.

We can ask whether it is possible to have find a stable ``ghost condensate''
vacuum for a theory with wrong-sign kinetic terms.
In fact, general relativity can be viewed as such a theory.
The {\it Vierbein} and spin connection form the gauge fields for
$ISO(3,1)$, the group of Lorentz transformations and translations on
4-dimensional Minkowski space.
The gauge group is noncompact, so fluctuations about the vacuum
$e^a_\mu = 0$ include ghosts, but fluctuations about
nontrivial vacua such as flat space $\avg{e^a_\mu} = \de^a_\mu$
are stable.
In our theory with a single scalar, we can imagine that
there are other terms such as $(\partial \phi)^4$ in
the lagrangian that ``stabilize'' the ghost in the same way the
$\phi^4$ term stabilizes the tachyon. One immediate difference
with the tachyon case is that all of these higher order terms are
non-renormalizable; if these terms are to become as important as
the $(\partial \phi)^2$ term as they must in order to stabilize
the theory, then it appears that a description in terms of
effective field theory is not possible. As we will see this is in
fact not the case, and there is a systematic low-energy expansion
for fluctuations around the ghost condensate. But let us press on
for a bit, and start by assuming that the lagrangian has the form
\beq
\scr{L} = M^4 P(X),
\eeq
where $X = \d^\mu \phi \d_\mu \phi$.
We have now chosen $\phi$ to have the dimension of length, so that
$X$ is dimensionless, while $M$ is a constant with the dimension of
mass. Note that we are neglecting terms involving more than one
derivative acting on $\phi$; we will come back to these terms
below. A typical form for the function $P$ we are considering is
shown in Fig. 4.

\begin{figure}
 \centering\leavevmode\epsfysize=8cm \epsfbox{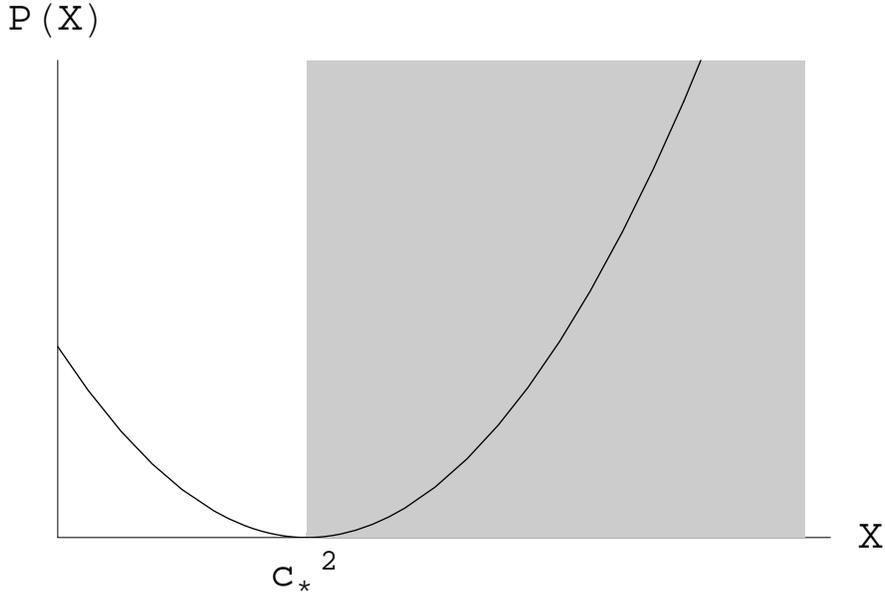}
 \caption{\label{fig:P}
The function $P$. The shaded region is the one for which the
small fluctuations about the background have healthy
two-derivative time and spatial kinetic terms.
 }
\end{figure}

The equations of motion that follow from this lagrangian are
\beq
\d_\mu \left[ P^\prime(X) \d^\mu \phi \right] = 0.
\eeq
{}From this
we see that we obtain a solution provided that $\d_\mu \phi =
\hbox{\rm constant}$. If $\d_\mu \phi$ is time-like, there is a
Lorentz frame where
\beq
\phi = c \, t ,
\eeq
where $c$ is a
dimensionless constant. These solutions are actually the maximally
symmetric solutions, since they break the $\phi$ shift symmetry
and the time translation symmetry down to an unbroken diagonal shift
symmetry.

Let us now consider small fluctuations about these solutions
\beq
\phi = c \, t + \pi.
\eeq
The lagrangian for
quadratic fluctuations is
\beq[piquadLeff]
\scr{L} = M^4 \left\{
\left[ P'(c^2) + 2c^2 P''(c^2) \right] \dot\pi^2 - P'(c^2)
(\nabla \pi)^2 \right\}.
\eeq
We see that the fluctuations
have the time and spatial kinetic terms of the usual signs
provided that $c$ is such that
\beq[pistab]
P'(c^2)> 0, \qquad P'(c^2) + 2c^2 P''(c^2) > 0.
\eeq
Around these
backgrounds the excitations of $\pi$ have positive energy, and so
any weak coupling of this sector to other sectors (such as gravity
and the standard model) will not suffer from the violent UV
instability associated with the ghosts around the usual background
with $\partial_\mu \phi = 0$. This does not uniquely fix $c$, any
value of $c$ satisfying these conditions is stable against small
fluctuations (see Fig. 4).

More generally, we have additional terms in the effective
lagrangian arising from terms with two or more
derivatives acting on $\phi$:
\beq[ghosteffLhigher]
\scr{L} = M^4
P(X) + M^2 S_1(X) (\Box \phi)^2 + M^2  S_2(X) \d^\mu \d^\nu \phi
\d_\mu \d_\nu \phi + \cdots\,.
\eeq
(We have assumed for
simplicity here that there is a $\phi \mapsto -\phi$ symmetry that
forbids terms with an odd number of $\phi$'s in the lagrangian).
 With these additional terms
included, there is still a solution for any value of $c$, since
the $\phi$ equation of motion always takes the form
\beq
\d_\mu
\bigl[ \hbox{\rm function\ of\ $\d \phi$, $\d^2 \phi$, $\ldots$} \bigr]
= 0.
\eeq
The quadratic lagrangian for $\pi$ now includes terms with more
derivatives on the $\pi$'s. We will see below that Hubble friction
drives us to a background where the spatial kinetic terms coming
from $P'$ vanish and that the leading spatial kinetic terms are
fourth order and arise from the $S_{1,2}$ terms in 
Eq.~(\ref{eq:ghosteffLhigher}).

We can now understand why it is possible to have a good low-energy
effective field theory for the physics, despite the appearance of
the non-renormalizable interactions $P(\partial \phi)^2,
S((\partial \phi)^2)$ and so on. The point is that the background
sits at some specific value of $c$, and therefore does not sample
large regions of the functions $P, S_{1,2}, \ldots\,$.
Terms with $n > 1$ derivatives acting on $\phi$ give
terms of order $\d^n \pi$, which are higher order in the derivative
expansion.
After Taylor expanding the coefficient
functions $P, S_{1,2}, \cdots$, we obtain
an effective lagrangian with a finite number of parameters that
controls small fluctuations about a given background.

Another way to look at this is to note that the background
is characterized by a symmetry breaking pattern.
For example, the ghost condensate breaks Lorentz
transformations down to spatial rotations. The $\pi$ field is
the Goldstone boson and must non-linearly realize the full
Poincare symmetry of the theory, and the $\pi$ effective
lagrangian is highly constrained by symmetry considerations.
This point of view will be more fully developed below.

\subsection{Hubble friction}

So far, we have a continuum of possible backgrounds labelled by
the parameter $c$. The situation changes in the presence of
gravity and an expanding universe. Gravity is coupled to the
effective lagrangian of the previous section via the usual minimal
coupling: 
\beq 
\scr{L} = \sqrt{-g} M^4 P(X) + \cdots, 
\eeq 
where
$X = g^{\mu\nu} \d_\mu \phi \d_\nu \phi$.
Again, we begin by ignoring the higher order terms $S_{1,2}$,
{\it etc\/}.
The energy-momentum
tensor of the ghost condensate is
\beq[ghostemtensor]
T_{\mu\nu} =
- M^4 P(X) g_{\mu\nu} + 2 M^4 P'(X) \d_\mu \phi \d_\nu \phi.
\eeq
In general, we expect the background to be 
an expanding FRW universe, 
with metric
\beq
ds^2 = dt^2 - a^2(t) d\Om^2,
\eeq where
$d\Om^2$ is the spatial metric for a maximally symmetric
3-dimensional space.
For $\phi$ depending only on $t$, the $\phi$
equation of motion in these coordinates is 
\beq[phirolling] 
\d_t
\left[ a^3 P'(\dot\phi^2) \dot\phi \right] = 0. 
\eeq 
Note that because this is the $\phi$ equation of motion, it holds 
even if some other sector is dominating the expansion of the
universe.

{}From here we see that
\begin{equation}
\dot{\phi} P(\dot{\phi}^2) = \frac{\mbox{const}}{a^3(t)} \, \to 0
\ \ \mbox{as} \ \  a \to \infty .
\end{equation}
So, in the far future as $a \to \infty$, we are driven to either
\begin{equation}
\dot{\phi} = 0
\quad\mbox{or}\quad
P^\prime(\dot{\phi}^2) = 0,
\end{equation}
depending on the initial condition.
The first option is what we are familiar with from massless scalar
fields in an expanding universe---the scalar velocity redshifts
to zero as the universe expands. However, the point with
$\dot{\phi} = 0$ is the one corresponding to the unstable theory
with the wrong sign kinetic term. 
The other possibility is that the theory is driven to
$\dot\phi = c_*$, with $P'(c_*^2) = 0$.
If $P''(c_*^2)>0$,
the fluctuations about this point are stable
with suitable higher derivative terms as we will see momentarily.

These conclusions are robust under inclusion of the higher-order
terms $S_{1,2}, \ldots\,$.
In the case where the universe becomes flat in the asymptotic
future,
$\dot{\phi} = c$ implies that  $\nabla_\mu \nabla_\nu \phi = 0$,
and so all of the higher order terms vanish in the equation of
motion and we are still driven to the point where
$P^\prime = 0$.
If the asymptotic metric is de Sitter space with a
Hubble constant $H$, it is no longer true that
for $\dot{\phi} = c$, all higher covariant derivatives vanish; for
instance, $\Box \phi = 3 H c$, so the new terms make
non-vanishing contributions to the equation of motion.
In this case, we are
not driven to $P'=0$ exactly, but the important point is that
Hubble friction singles out {\it some} value of $c_*$ with 
$P'(c_*^2)\sim$ powers of $H/M$.
These are small for  $H/M \ll 1$, which is required by validity
of the low-energy effective theory.

\subsection{Low energy dispersion relation}

We have seen that gravity forces us into an extremum
$P^\prime(c_*^2) = 0$ (for an asymptotically flat spacetime).
Recall that the spatial kinetic terms for $\pi$ from the expansion
of $P$ about the background are proportional to $P^\prime$, so if
$P^\prime = 0$ there is no $(\nabla \pi)^2$ term in the quadratic
lagrangian for $\pi$. However, the fluctuations can still be
stabilized by $ (\nabla^2 \pi)^2$ terms, which arise from the
higher order terms proportional to $S_{1,2}$, \beq[k4terms] \De
\scr{L} = M^2 \left[ S_1(c_*^2) + S_2(c_*^2) \right] ( \nabla^2
\pi) ^2 + \cdots. \eeq This stabilizes fluctuations about $\phi =
c_* t$ provided that $S_1(c_*^2) + S_2(c_*^2) < 0$. The quadratic
lagrangian for $\pi$ is then
\begin{equation}
{\cal L} = \frac{1}{2} M^4\dot{\pi}^2 - \frac{1}{2}\bar{M}^2
(\nabla^2 \pi)^2,
\end{equation}
where we have put $P^{\prime \prime}(c_*^2) = \sfrac 14$, which can always be
achieved by rescaling between $P$ and $M^4$, and we have
defined $\bar{M}^2 = - 2 M^2 (S_1(c_*)^2 + S_2(c_*^2))$. The $\pi$
modes then have the low-energy dispersion relation
\begin{equation}
\om^2 = \frac{\bar{M}^2}{M^4} k^4.
\end{equation}


This conclusion also holds for
an asymptotically de Sitter space. 
As an example, consider the lagrangian of the form
\begin{equation}
{\cal L}=\sqrt{-g}M^4 \left[ P(X)+Q(X)R(\Box \phi)\right].
\end{equation}
For $\phi$ depending only on $t$, the equation of motion 
(\ref{eq:phirolling}) is modified to
\begin{equation}
\d_t
\left\{ a^3 \left[\left(P'(\dot\phi^2)+Q'(\dot\phi^2)R(\ddot\phi+3H\dot\phi)) 
2 \dot\phi - \d_t (Q(\dot\phi^2)R'(\ddot\phi+3H\dot\phi)\right)\right]
\right\} = 0. 
\end{equation}
As $a\to \infty$ we have
\beq
\left[ P'(\dot\phi^2)+Q'(\dot\phi^2)R(\ddot\phi+3H\dot\phi)\right]
2 \dot\phi &\to 0 \\
Q(\dot\phi^2)R'(\ddot\phi+3H\dot\phi) &\to \mbox{const.}
\eeq
Expanding in powers of $\pi$, we find that
the coefficient of the $(\nabla \pi)^2$ term is exactly
$-M^4(P'(\dot\phi^2)+Q'(\dot\phi^2)R(\ddot\phi+3H\dot\phi))$, which
is driven to zero (for $\dot\phi \neq 0$) by the expansion of
the universe. The generality of the $k^4$ dispersion relation
is easier to see from the non-linear realization formulation,
to be discussed in section 6.

If the universe did not start in the special configuration $\dot\phi = c_*$, 
the $(\nabla \pi)^2$ term will not vanish exactly today and its
size depends on the history of the universe. As we will see later,
even in the absence of the $(\nabla \pi)^2$ term
there will be a $k^2$ piece induced by mixing with gravity in the
dispersion relation.
The coefficient of the $(\nabla \pi)^2$ term can be made negligibly
small compared with the gravity-induced piece 
if it was driven to zero \eg\ by a period of inflation in the early universe. 
On the other hand,
it would also be interesting to study the effects of a significant 
$(\nabla \pi)^2$ term that may still remain in the current universe.

\subsection{Two kinds of energy and (anti)gravity}

Our background
\begin{equation}
\langle \phi \rangle = c_* t
\end{equation}
spontaneously breaks time translation invariance. It also breaks
the shift symmetry $\phi \to \phi + c$. However, a diagonal sum of
these two shift symmetries is left unbroken as the time
translation in the background. The combination
\begin{equation}
{\cal E}_{\rm pp} = \myint d^3 x\, T_{00} - c_* Q
\end{equation}
of the Noether charge $\int \!d^3 x \, T_{00}$
associated with time translations,
and the charge $Q$ associated with the $\phi$ shift symmetry,
is the 
conserved energy in this background. The usefulness of
${\cal E}_{\rm pp}$ is that
it is the conserved energy that an experimentalist would infer
from the dynamics of the $\pi$ fluctuations
\begin{equation}
{\cal E}_{\rm pp} =
\myint d^3 x \left[ \frac{1}{2} M^4 \dot{\pi}^2 + \frac{1}{2} \bar{M}^2
(\nabla^2 \pi)^2 + \cdots \right].
\end{equation}
It is the conservation of this quantity that ensures the stability
of the vacuum against small fluctuations. On the other hand, this
is {\it not} the energy $T_{00}$ that couples to gravity. Indeed,
from the expression for $T_{\mu \nu}$ we can infer that
\begin{equation}
T_{00} = c_*^3 M^4 \dot{\pi} + \cdots.
\end{equation}
This begins at {\it linear} order in $\dot{\pi}$. Therefore,
$T_{00}$ can be either positive or negative, depending on the sign
of $\dot\pi$, and localized excitations of the ghost condensate
can give rise to antigravity! Together with the unusual dispersion
relation for $\pi$, this will  be the root of the exotic
infrared phenomena we will find below.

\section{de Sitter expansion from ghost condensation}

Our background spontaneously breaks Lorentz invariance, since
there is a preferred frame in which $\phi$ is spatially isotropic.
This is not conceptually any different from the Lorentz violation
due to any cosmological fluid that fills the universe, like
the cosmic microwave background radiation. Indeed, because of
Hubble friction it is natural that this frame coincides with the
frame in which the cosmic microwave background is isotropic. There
is however an important difference between the ghost condensate
and conventional forms of matter and radiation: it does not
redshift away as the universe expands. Indeed, in the far future
the energy-momentum tensor becomes
\beq
T_{\mu\nu} \to -g_{\mu\nu}
M^4 P(c_*^2),
\eeq
\ie, precisely the form of a cosmological
constant! If $P(c_*^2) < 0$, the universe will approach a de
Sitter metric. However the ghost condensate is {\it not} a
cosmological constant, it is a physical fluid with a physical
scalar excitation.

The cosmological constant in this theory is simply $\Lambda = -
M^4 P(0)/\MP^2$. Even if this vanishes, the ghost condensate can
drive the current acceleration of the universe if
$M^4 P(c_*^2) \sim -(10^{-3}\eV)^4$, and assuming that there are no very large or
small dimensionless parameters in $c$, this implies that $M \sim
10^{-3}\eV$. It is also possible for the ghost condensate to drive
an early inflationary phase of the universe, as will be discussed
in \Ref{ghostinfl}.
The fact that accelerated expansion of the universe can be driven
by the kinetic terms was discussed before, under the names of
``k-inflation'' \cite{Armendariz-Picon:1999rj} and
``k-essence'' \cite{Armendariz-Picon:2000dh}.
In those works, $\dot\phi$ is required to move over a large range of
the kinetic function, which is truncated to low orders or assumed
to be of some special form. These models therefore require some
knowledge
of the fundamental theory and are not based on consistent
effective field theories.

If $\phi$ is not exactly at the solution $\dot\phi = c_*$ where
$P'(c_*^2) = 0$, then we have
\beq[ghostmatt]
P'(\dot\phi^2) \sim \frac{1}{a^3}.
\eeq
{}From the energy-momentum tensor of the ghost
condensate \Eq{ghostemtensor}, we see that a nonzero $P'$ couples
to gravity like non-relativistic matter, namely $T_{00} \sim M^4
P'$, $T_{ij} = 0$. Therefore in the approach to its final position
at $\dot{\phi} = c_*$, the ghost condensate sources gravity like
non-relativistic matter. This raises the prospect that the ghost
condensate may contribute to dark matter as well as
the vacuum energy,
hinting that these phenomena may both
have their origin in a modification of gravity at large distance
scales. This possibility will be investigated in future work.

\section{Low-energy effective theory}

We would now like to examine the low-energy effective theory for
$\pi$ in more detail, in the absence of gravity.
We will consider
the theory expanded around the background $P^\prime(c_*^2) = 0$,
which we are driven to by Hubble friction.
We show that this theory is truly healthy in the absence of
gravity, in contrast to the analogous scalar sectors in DGP
and massive gravity, which only become healthy due to the coupling
to gravity.
In particular, we will address the regime of validity of the effective
theory and the question of stability.

Expanding $X = c_*^2 + 2 c_* \dot{\pi} + \dot{\pi}^2 - (\nabla
\pi)^2$, we can see that the low-energy effective action for the
$\pi$'s begins as
\beq
S = \myint d^3 x\, dt \Bigl[ & \frac{1}{2} M^4 \dot{\pi}^2 - \frac{1}{2}
\bar{M}^2 (\nabla^2 \pi)^2 \\
& - \frac{M^4}{2c_*} \dot{\pi} (\nabla
\pi)^2 + \frac{M^4}{8c_*^2}(\nabla \pi)^4  + \cdots \Bigr],
\eeq
where we have suppressed the higher order terms that arise from
the $S_{1,2}$ terms giving rise to the $(\nabla^2 \pi)^2$
terms.

Clearly for this to be a sensible low-energy effective theory, the
cutoff has to be beneath the scales $M, \bar{M}$, which we will
take to be roughly comparable.
If we canonically normalize the $\dot\pi^2$ kinetic term,
we find that the $(\nabla^2 \pi)^2$ term and the cubic
interaction $\dot{\pi} (\nabla \pi)^2$ term are both mass dimension 6
operators.
Which term is more important at low energies?
A closely related question is that of stability.
The (particle physics) energy is independent of the
$\dot{\pi} (\nabla \pi)^2$ term
because it is linear in $\dot\pi$, but the quartic term
$(\nabla \pi)^4$ contributes a negative-definite term to the
energy density that will be larger than the positive contribution
from the $(\nabla^2 \pi)^2$ term provided that
$\nabla\pi \gsim k/M$.
It is therefore possible to find directions in field configuration
space with arbitrarily long wavelengths in which the energy
can be made more and more negative.
This means that we have to address the nonlinear stability of the
theory.
This is most pressing at quantum level:
what stops quantum fluctuations from lowering the
energy and leading to vacuum instability?

The key to understanding all of these issues is to do the proper
power-counting analysis for the scaling dimension of all the
operators in our effective theory (see \Ref{Polchinski:1992ed}
for a very clear discussion of these issues).
In a relativistic
theory, the scaling and mass dimension of couplings are the same,
while this is not the case in the non-relativistic theory we are
studying. Let us begin then by identifying the scaling dimensions
as the energy is scaled by $E \mapsto s E$, where $s$ is some
constant, or in terms of time $t \mapsto s^{-1} t$. We determine the
scaling dimension of space $x$ and the field $\pi$ by requiring
that the quadratic part of the action is invariant. This then
fixes
\beq
E &\mapsto s E, \\
t &\mapsto s^{-1} t, \\
x &\mapsto s^{-1/2} x, \\
\pi &\mapsto s^{1/4} \pi.
\eeq
The scaling of $x$ is expected since $ \omega \propto k^2$, as
$\omega \mapsto s \omega$ we should take $k \to s^{1/2} k$ and so $x
\mapsto s^{-1/2} x$. With this scaling, it is easy to check that the
leading interaction
\begin{equation}
\myint d^3 x\, d t \, M^4 \dot{\pi} (\nabla \pi)^2
\end{equation}
scales as $s^{1/4}$. It is therefore (just barely!)\ an irrelevant
operator, becoming less and less important at low energies
compared to $M$. All the other operators are even more irrelevant
than this leading one. 
This implies that there is a regime of low energies and momenta,
and low field amplitudes, where the expansion is under control.%
\footnote{This is a non-trivial result. For instance,
suppose it had turned out that the leading spatial kinetic term
for $\pi$ was of the form $(\nabla \nabla^2 \pi)^2$, then the
scaling dimensions of the field would have been different ($x \mapsto
s^{-1/3} x$ and $\pi \to s^0 \pi$), and the leading operator
$\dot{\pi} (\nabla \pi)^2$ would have scaled as $s^{-1/3}$
and been {\it relevant}, becoming
strong and leading to a breakdown on the effective theory at low
energies. Fortunately this did not happen for us. It is amusing to
note that our leading operator is more irrelevant for all
spacetime dimensions $d > 4$, while for $d=3$ it is classically
marginal! In this case the 1-loop beta function must be computed
to determine the IR behavior of the theory.}
In particular, the quantum fluctuations are not large enough to
make the higher order terms compete with the quadratic terms.
There are therefore no large quantum instabilities in the IR;
only in the UV where the theory becomes strongly coupled is there
a danger of instabilities from the higher order terms.
This is however sensitive to the UV completion, and in this respect the situation
is similar to QED or any other theory that gets strongly coupled
in the UV.
In any case, if we work with a UV cutoff $\Lambda$ that
is somewhat smaller than the scales $M,\bar{M}$, then the strongest
dimensionless coupling strength in the theory is set by powers of
$\lambda = \Lambda/M$. The timescale for any quantum instability,
if it exists, will be exponentially small in $1/\lambda$, and can
easily be longer than the age of the universe.

For completeness note also that there is an operator that is not
forbidden by any symmetries, namely $(\nabla \pi)^2$.
This is definitely
a relevant operator, however, as we have seen, Hubble friction
always drives the background to a point where the coefficient of
this operator vanishes.

The fact that the theory is healthy independently of gravity is
in contrast with case of the massive gravity or the DGP model.
In our model there is a limit where the scalar decouples, while
keeping the strength of gravity fixed.
To do this, we take $\MP \to \infty$ and also all matter sources
$T_{\mu\nu} \to \infty$, keeping the gravitational scale
$T_{\mu\nu} / \MP^2$ fixed, while also keeping $M$ fixed.
In massive gravity or DGP model, there is a scalar
coupled to ordinary matter with the same strength as the graviton
and its effects do not vanish in this limit.

The scales $M$ that we will be considering will turn out to be
quite a bit smaller than the TeV scale, in the range $10^{-3}$ eV
$\lsim M \lsim $ 10 MeV, and the $\pi$ sector
will need to be embedded in a UV
completion above this scale. However, gravity and the standard
model are very weakly coupled to the $\pi$ sector, 
so the breakdown of the $\pi$ effective theory at $M$
will not appreciably affect the standard model sector.

\section{Direct coupling to Standard Model fields}
The standard model fields need not have any direct couplings to
the ghost sector (other than indirectly through gravity). However,
it is interesting to consider possible direct couplings of the
standard model fields to $\phi$, preserving the shift symmetry on
$\phi$.
The cutoff of the standard model sector can be much larger
than the cutoff of the ghost sector as long as the couplings
between them are weak enough.
The leading derivative coupling of $\phi$ to standard model fermions
is familiar from Goldstone boson interactions, and is of the form
\begin{equation}
\De\scr{L} = \sum_\psi\frac{ c_\psi }{F} \bar{\psi}
\bar{\sigma}^\mu \psi
\partial_\mu \phi.
\end{equation}
(Here we have reverted to canonical normalization for the $\phi$
field.)
A systematic discussion of all possible Lorentz violating effects
for standard model fields in flat space can be found in
\Ref{Colladay:1996iz,Coleman:1998ti}.
Note that these couplings can be removed by field
redefinitions on the fermions $\psi \to e^{i c_\psi \phi/F} \psi$;
however, if the symmetry $\psi \to e^{i c_\psi \theta} \psi$ is
broken by mass terms or other couplings in the lagrangian, the
interaction cannot be removed. If $\psi, \psi^c$ are paired with
a non-zero Dirac mass term $m_D \psi \psi^c$, then the vector
couplings (with $c_\psi + c_{\psi^c} = 0$) can be
removed, while the axial couplings
\begin{equation}
\Delta {\cal L} \sim \frac{1}{F} \bar{\Psi} \gamma^\mu \gamma^5
\Psi
\partial_\mu \phi
\end{equation}
remain. Expanding $\phi = M^2 t + \pi$, we have
\begin{equation}
\De\scr{L} \sim \mu \bar{\Psi} \gamma^0 \gamma^5 \Psi + \frac{1}{F}
\bar{\Psi} \gamma^\mu \gamma^5 \Psi \partial_\mu \pi, \, \,
\mbox{where} \, \, \mu \equiv \frac{M^2}{F}
\end{equation}
The first term is a Lorentz and CPT violating term, that gives
rise to a different dispersion relation for particles and their
antiparticles. In particular, a left helicity particle and right
helicity antiparticle have dispersion relations of the form
\begin{equation}
\omega = \sqrt{(|p| \pm \mu)^2 + m_D^2}
\end{equation}
where the $+$ sign is for the left-helicity particle and the $-$
sign is for the right-helicity antiparticle, and the signs are
reversed for a right-helicity particle and left-helicity
antiparticle~\cite{Kostelecky:1999zh,Andrianov:2001zj}.
Furthermore, if the earth is moving with respect to
the background in which the condensate is spatially isotropic,
there is also an induced Lorentz and CPT-violating mass term of
the form
\begin{equation}
\mu \bar{\Psi} \vec{\gamma} \gamma^5 \Psi \cdot \vec{v}_{\rm earth}.
\end{equation}
In the non-relativistic limit, this gives rise to an interaction
Hamiltonian
\begin{equation}
\mu \vec{S} \cdot \vec{v}_{\rm earth}.
\end{equation}
The experimental limit on $\mu$ for the coupling to the electrons
is of the order $\mu \lsim 10^{-25}$ GeV~\cite{Heckel:1999sy}, and
to the proton and the neutron are
$\lsim 10^{-24}$~GeV~\cite{Phillips:2000dr,Cane:2003wp}, assuming
$|\vec{v}_{\rm earth}| \sim 10^{-3}$.
A review of Lorentz and
CPT experimental tests and extensive references can be found
in~\Ref{Bluhm:2003ne}.

Also strikingly, the exchange of $\pi$ gives rise to a new
long-range force. In the non-relativistic limit, we have a
derivative coupling to spin
\begin{equation}
\De\scr{L} \sim \frac{1}{F} \vec{S} \cdot \nabla \pi,
\end{equation}
As with usual Goldstone bosons, the exchange of $\pi$ produces a
long-range spin-dependent potential. Ordinarily for Goldstone
bosons this produces a $1/r^3$ potential, but here, because of the
$k^4$ dispersion relation instead of the usual $k^2$ dispersion
relation, we get a $1/r$ potential:
\begin{equation}
V \sim \frac{M^4}{\tilde{M}^2 F^2} \frac{\vec{S}_1 \cdot
\vec{S}_2 - 3 (\vec{S}_1 \cdot
\hat{r}) (\vec{S}_2 \cdot \hat{r}) }{r}.
\end{equation}
Therefore these goldstones mediate spin-dependent inverse-square
law forces! Note that we have imagined exactly static sources in
deriving this force law, ignoring the retardation effects---the
$\omega^2$ piece in the denominator of the $\pi$ propagator. For
massless particle exchange in relativistic theories, retardation
effects can be ignored as long as the sources are static on a
timescale longer than the time it takes light to travel between
them. In our case, however, since $\omega^2 \propto k^4$, for
sources a distance $r$ apart the static limit force law we have
found is only valid on timescales longer than
\begin{equation}
\tau \sim \omega^{-1} \sim {M}\, r^2.
\end{equation}

We have focused here on direct couplings to the $\phi$ field that
are linear in $\phi$. These can all clearly be forbidden by a
$\phi \to - \phi$ symmetry. There are however some couplings
between the standard model fields and $\phi$ that are inevitably
generated by graviton loops. The leading operators of this type
have the form
\begin{equation}
\frac{1}{\MP^4} {\cal O}_{\rm SM}^{\mu \nu} \partial_\mu \phi
\partial_\nu \phi,
\end{equation}
where ${\cal O}_{\rm SM}^{\mu \nu}$ is some dimension $4$ standard
model operator. Putting in the background $\dot{\phi} = M^2$ then
gives rise to tiny dimensionless Lorentz violating effects of size
$\sim M^4/\MP^4$, for instance differing speeds of light for
different particle species. Also, for ${\cal O}_{\rm SM}^{\mu \nu} =
T^{\mu \nu}$, this operator can act as a coherent source for
$\phi$, however because of the $\MP$ suppression, this is never
competitive with the existing terms in the lagrangian for $\phi$.

There is a large literature on Lorentz violation in the standard
model~\cite{CPT:2002}. 
The rotationally invariant effects have been thought of as
arising from a non-zero value for the time component of a
background vector field $A_\mu$, $\langle A_0 \rangle \neq 0$. Our
model is of this type with $A_\mu \to \partial_\mu \phi$, and
therefore provides a fully consistent framework to study Lorentz
violation, including gravitational effects.

\section{Ghost condensate coupled to gravity}
In this section, we construct the effective theory of the ghost
condensate coupled to gravity. We can simply do this by minimally
coupling our lagrangian for $\phi$ to gravity, but it also
instructive to see how this works directly in a ``unitary
gauge," where the modification of gravity is seen transparently.

To define unitary gauge, note that we can use general coordinate
invariance to use $\phi$ as a time coordinate, \ie, we choose
coordinates $(t, x^i)$ such that
\beq
\phi(t, x) = t.
\eeq
This is
a good gauge choice for small fluctuations about a time-dependent
solution such as the one we are considering. This choice of gauge
eliminates $\phi$ as a degree of freedom, and leaves residual
gauge freedom corresponding to time-dependent spatial
diffeomorphisms: \beq[residdiff] t \mapsto t, \qquad x^i \mapsto
x^{\prime i}(t, x). \eeq We can write the unitary gauge lagrangian
in terms of quantities that transform simply under these residual
diffeomorphism symmetry.

Let us first see this works at quadratic level in
the action expanding around flat space, $g_{\mu \nu} = \eta_{\mu
\nu} + h_{\mu \nu}$. Under general diffeomorphisms generated by
$x^\mu \to x^\mu + \xi^\mu(x)$, we have as usual
\begin{equation}
\delta h_{\mu \nu} = -(\partial_\mu \xi_\nu + \partial_\nu
\xi_\mu).
\end{equation}
Under the residual unbroken diffeomorphisms generated by
$\xi_i$, we then have
\begin{equation}
\delta h_{00} = 0, \quad \delta h_{0 i} = -\partial_0 \xi_i , \quad
\delta h_{ij} = -(\partial_i \xi_j + \partial_j \xi_i).
\end{equation}
What invariants can we use to build the action? Recall that we are
assuming that flat space is a good background here, so any action
must begin at quadratic order in the $h$'s. Clearly, since
$h_{00}$ is invariant, the leading term is of the form
\begin{equation}
\myint d^3 x \, dt\: \frac{1}{8} M^4 h_{00}^2.
\end{equation}
As we have mentioned, because these terms do not preserve the full
diffeomorphism invariance of the theory, we are really introducing
an additional degree of freedom. To see this explicitly, it is
convenient to re-introduce the field $\pi$ that restores the full
diffeomorphism symmetry of the theory---we can achieve this simply
by performing a broken $\xi^0$ diffeomorphism and promoting
$\xi^0 = \pi$ to a field.
Then
\begin{equation}
h_{00} \to h_{00} - 2 \partial_0 \pi, \quad h_{0i} \to h_{0i} -
\partial_i \pi, \quad h_{ij} \to h_{ij},
\end{equation}
and we see that the $M^4 h_{00}^2$ action has generated a time
kinetic term $\frac{1}{2}M^4 \dot{\pi}^2$ but no spatial kinetic
term for $\pi$. In fact, it is easy to see that only adding the
$M^4 h_{00}^2$ term to the action, we have not modified GR at all
(at least classically), since this is a partial
gauge-fixing term that
fixes to $h_{00} = 0$ gauge. We must therefore go to higher order,
to make $\pi$ a dynamical field, and see a real modification of GR
rather than just a gauge fixing.%
\footnote{The $h_{00}^2$ term does not fix time reparameterization
globally.
In particular, static solutions are not unique.
The correct procedure is to define static solutions as the limit of
time-dependent solutions, as will be done
in specific examples below.}

In this language, it is easy to see why we do not get a $(\nabla\pi)^2$
kinetic term.
We could get this from terms of the form
$h_{0i}^2$ in the action, since upon introducing the $\pi$ we have
$h_{0i} \to h_{0i} -
\partial_i \pi$. However, $h_{0i}^2$ is not invariant under the
residual diffeomorphisms! There are invariants that can be
constructed, but they involve higher derivatives, for instance
\beq
K_{ij} &= \frac12 (\partial_0 h_{ij}- \partial_j h_{0i} -
\partial_i h_{0j})
\nonumber\\
&\to \frac12 (\partial_0 h_{ij}- \partial_j h_{0i} -
\partial_i h_{0j} + \d_i \d_j \pi)
\eeq
is invariant (and is the linearized extrinsic curvature of
constant time surfaces in the theory). The leading terms at
quadratic level in the effective theory are of the form
\begin{equation}
\label{eq:k_squared}
S = \myint d^3 x\, dt  \left[ -\frac{1}{2}\tilde{M}^2 K_{ii}^2 - \frac{1}{2}
\tilde{M}^{\prime 2} K_{ij} K_{ij} \right].
\end{equation}
Terms linear in $K$ can be forbidden by time-reversal invariance.
If we do add for example $h_{00}K \to \dot{\pi}\nabla^2 \pi + \cdots$,
it contributes a term $\omega k^2$
to the $\pi$ dispersion relation. However, this does not qualitatively
change the physics as $\omega \sim k^2$ still holds. Terms linear
in $\dot{\pi}$ do not contribute to the Hamiltonian and hence
do not affect stability either. 

Upon re-introducing $\pi$, (\ref{eq:k_squared})
gives us the $k^4$ spatial kinetic terms,
\begin{equation}
S = \myint d^3x \, dt \left[  -\frac12 \bar{M}^2 (\nabla^2 \pi)^2 
+ \cdots \right],
\qquad
\bar{M}^2 \equiv (\tilde{M}^2 + \tilde{M}^{\prime 2}),
\end{equation}
so we see that the $k^4$ spatial kinetic terms are a direct
consequence of the assumption that flat space is a good
background, and the unbroken residual spatial diffeomorphisms of
the theory.

We can repeat the same analysis for a de Sitter background, the
metric is of the form
\begin{equation}
d s^2 = ( 1 + h_{00}) d t^2 - e^{2 H t} dx^i dx^j (\delta_{ij} -
h_{ij}) +2 h_{0 i} dt dx^i,
\end{equation}
and the linearized diffeomorphism transformations are
\begin{equation}
\delta h_{00} = -2 \partial_0 \xi_0, \, \, \delta h_{0 i} =
-\partial_i \xi_0 - e^{2 H t} \partial_0 \xi_i, \, \, \delta
h_{ij} = 2 H \xi_0 \delta_{ij} - \partial_i \xi_j - \partial_j
\xi_i,
\end{equation}
where $\xi_\mu = \eta_{\mu\nu} \xi^\nu$.
As before, we can restore full diffeomorphism invariance by
promoting $\xi_0$ to a dynamical field $\pi$.
Then the only invariants under linearized diffeomorphisms are
\beq
h_{00} - 2 \dot\pi,
\quad\mbox{and}\quad
\dot{h}_{ij} - e^{-2Ht} (\d_i h_{0j} + \d_j h_{0i} - \d_i \d_j \pi)
+ 2 \de_{ij} H \dot\pi.
\eeq
As before, we cannot write an invariant term that contains a
$(\nabla \pi)^2$ kinetic term.


Let us now proceed to the systematic construction of the effective
theory at non-linear level. First, consider the scalar quantity
\beq X = g^{\mu\nu} \d_\mu \phi \d_\nu \phi. \eeq In unitary
gauge, \beq X \to g^{00}, \eeq and we see that in unitary gauge
the lagrangian can be an arbitrary function of $g^{00}$. Writing
the lagrangian in terms of $X$ is useful because it is obvious how
to generalize to an arbitrary gauge.

We can also define the unit vector perpendicular to the surfaces of constant
$\phi$:
\beq
n_\mu = \frac{\d_\mu \phi}{\sqrt{X}} \to \frac{\de_\mu{}^0}{\sqrt{g^{00}}}.
\eeq
This can be used to project any tensor into tensors with all indices parallel
to the surfaces of constant $\phi$.
For example, the induced metric is simply
\beq
\ga_{\mu\nu} = -(g_{\mu\nu} - n_\mu n_\nu),
\eeq
which satisfies $n^\mu \ga_{\mu\nu} = 0$, and is therefore a metric
on the surfaces of constant $\phi$.
We can use this to write other tensors such as the
intrinsic curvature $R(\ga)$ and the extrinsic curvature
\beq
K_{\mu\nu} = \ga_\mu{}^\rho \nabla_\rho n_\nu,
\eeq
which is a symmetric tensor satisfying $n^\mu K_{\mu\nu} = 0$.

Since the residual diffeomorphisms depend on time, $\d_0$ is not a
covariant derivative.
The covariant time derivative on tensors with all indices
parallel to surfaces of constant $\phi$ is
\beq
D^\phi T^{\cdots}{}_{\cdots} = \frac{1}{\sqrt{X}}
(n^\mu \nabla_\mu T^{\cdots}{}_{\cdots})^\parallel,
\eeq
where `$\parallel$' denotes the projection onto surfaces of constant
$\phi$ using $n_\mu$.
(The normalization is dictated by the fact that $n^\mu \d_\mu = \d^0$
in unitary gauge.)

The most general effective lagrangian invariant under spatial
diffeomorphisms in unitary gauge is therefore
\beq[LeffUnitaryGauge] \bal \scr{L} = \sqrt{\ga} \bigl[ & F(X) +
\sfrac 12 Q_1(X) K^2 + \sfrac 12 Q_2(X) K^{ij} K_{ij} + \cdots
\\
& \quad + \hbox{\rm terms\ involving\ time\ derivatives} \bigr].
\eal\eeq Here the $ij$ indices are raised and lowered with
$\gamma_{ij}, \gamma^{ij}$.  Again we have assumed the
time-reversal symmetry forbidding the terms linear in $K$. Note
also that at this order we could have written down a term
involving the intrinsic curvature $R^{(3)}$, however, the full
four-dimensional curvature $R^{(4)}$ is given by $R^{(4)} =
R^{(3)} + K_{ij}^2 - K^2$, so we can always write a term linear in
$R^{(3)}$ as a usual Einstein-Hilbert term plus terms proportional
to $K^2, K_{ij}^2$.

Note  that we can easily restore the dependence on the ghost
fluctuations $\pi$ by using the general formulas above. Using the
relation $\sqrt{-g} = \sqrt{\ga} / \sqrt{X}$, we see that \beq
F(X) = M^4 \frac{P(X)}{\sqrt{X}}. \eeq

Let us now consider the expansion about flat space in unitary
gauge. The terms that do not involve derivatives in the unitary
gauge lagrangian are \beq \scr{L} = \frac{\sqrt{\ga} M^4
P(X)}{\sqrt{X}}. \eeq We would like to have a solution where
$g_{\mu\nu} = \eta_{\mu\nu}$, so $X = \eta^{00} = 1$. It is easy
to see that this requires $P(1) = 0$ to avoid tadpoles for
fluctuations of the spatial metric, and $P'(1) = 0$ to avoid
tadpoles for $g^{00}$. The fact that $P(1) = 0$ can be interpreted
as the condition that there is no cosmological constant (see
\Eq{ghostemtensor}). The fact that $P'(1) = 0$ implies that the
ghost fluctuations have no $k^2$ term in their dispersion
relation as we have seen, and the leading terms are the $k^4$
ones (see \Eq{piquadLeff}).

Summarizing, the leading low-energy effective action is, in
unitary gauge,
\begin{equation}
S = \myint d^3x\, dt \sqrt{\gamma} \left[\frac{1}{8} M^4 (X -
1)^2 \, - \, \frac{1}{2} \tilde{M}^2 K^2 \, - \, \frac{1}{2}
\tilde{M}^{\prime 2} K_{ij} K^{ij} + \cdots \right].
\end{equation}

\section{Infrared modification of gravity}

We now consider the modification of gravity in the ghost
condensate, working at linearized level. At this order, the only
effect comes from the mixing between $\pi$ and gravity, and so it
suffices to only consider scalar perturbations of the metric. A
general scalar perturbation can be parameterized by
\begin{equation}
h_{00} = 2\Phi^\prime, \quad
h_{0i} = \partial_i \chi, \quad
 h_{ij} =
2 \delta_{ij} \Psi^\prime + \partial_i \partial_j \eta.
\end{equation}
Using the the diffeomorphisms generated by $\xi_i =\frac12
\partial_i \eta,\, \xi_0=\chi-\frac12 \partial_0 \eta$,
we can bring the metric to the longitudinal gauge,
\begin{equation}
h_{00} = 2 \Phi, \quad
 h_{0i} = 0, \quad
  h_{ij} = 2\Psi \delta_{ij}.
\end{equation}

We can analyze the physics at the level of the quadratic action
(or equivalently the linearized equation of motion). Let us begin
by working at the level of the action. The Einstein-Hilbert action
in this gauge is,
\begin{equation}
S_{\rm EH} = \MP^2 \left[-3 (\partial_0 \Psi)^2 - 2 \nabla \Psi
\cdot \nabla \Phi + (\nabla \Psi)^2 \right], \quad
\MP^2 \equiv \frac{1}{8\pi G}.
\end{equation}
Note that despite the appearance of the time derivative on $\Psi$
in this action, neither $\Phi$ nor $\Psi$ are dynamical fields.
Indeed it is easy to go to momentum space and diagonalize the $2
\times 2$ kinetic matrix mixing $\Phi,\Psi$; the zeros of the
determinant of this matrix give the dispersion relation which is
just $k^4 = 0$. This is in keeping with the fact that the only
propagating fields in gravity are the two polarizations of the
spin-two massless gravitons. Note also that in the Newtonian
limit, $\omega^2 \ll k^2$, so the action reduces to $-2 \nabla
\Psi \cdot \nabla \Phi + (\nabla \Psi)^2$, and the $\Psi$ equation
of motion then fixes $\Psi = \Phi$.

We now include the ghost condensate.
We again work in the Newtonian limit where $\omega^2 \ll k^2$.
Then at quadratic level the lagrangian for $\pi$ becomes
\begin{equation}
{\cal L}
= \frac{1}{2} M^4 (\Phi-\dot{\pi})^2 - \frac{1}{2} \bar{M}^2
(\nabla^2 \pi)^2.
\end{equation}
We can then put $\Psi$ to its equation of motion $\Psi = \Phi$,
and go to canonical normalization $\Phi = \Phi_c/(\sqrt{2}\MP),
\pi = \pi_c/M^2$, to find the lagrangian
\begin{equation}
{\cal L} = -\frac12 (\nabla \Phi_c)^2 + \frac{1}{2} \left(
\frac{M^2}{\sqrt{2}\MP} \Phi_c -\dot{\pi}_c \right)^2 -
\frac{\bar{M}^2}{2M^4} (\nabla^2
\pi_c)^2.
\end{equation}
We can see that there are small mixing terms between $\Phi_c$ and
$\dot{\pi}_c$. Going to momentum space, the lagrangian in the $(\pi_c,
\Phi_c)$ basis is
\begin{equation}
{\cal L} = \frac12 (\pi_c \, \Phi_c) \left(
\begin{array}{cc} \omega^2 - \alpha^2 k^4/ M^2 & - i m
\omega \\ i m \omega & -k^2 + m^2
\end{array} \right) \left( \begin{array}{c} \pi_c \\ \Phi_c
\end{array} \right),
\end{equation}
where we have defined
\begin{equation}
m \equiv \frac{M^2}{\sqrt{2}\MP}, \quad \alpha^2 \equiv
\frac{\bar{M}^2}{M^2}.
\end{equation}

We see that the $\pi$ dispersion relation is
modified from $\omega^2 = \alpha^2 k^4/M^2$ due to mixing with
gravity. 
The dispersion relation is found by setting the
determinant of the $2 \times 2$ matrix above to zero,
which gives
\beq[ghostJeans]
\omega^2 = \alpha^2 \frac{k^4}{M^2} - \alpha^2 \frac{m^2}{M^2}
k^2 = \alpha^2 \frac{k^4}{M^2} -
\frac{\alpha^2 M^2}{2\MP^2}
k^2.
\eeq
For very low momenta $k < m$, 
we find $\om^2 < 0$, signaling an instability.
This is the analog for our fluid
of the usual Jeans instability for ordinary fluids; for a fluid of
mean density $\rho$ and pressure $p$, the usual Jeans instability
shows up as a modification of the dispersion of the form%
\footnote{Strictly speaking, the dispersion relation \Eq{Jeans}
is not correct, because in ordinary gravity flat space is not a solution 
when $\rho \ne 0$ (the `Jeans swindle').
In the ghost condensate, flat space \emph{is} a solution, so the
dispersion relation \Eq{ghostJeans} is not a swindle.}
\beq[Jeans]
\omega^2 = \frac{p}{\rho} k^2 - \omega_J^2, \, \mbox{where} \, \,
\omega_J^2 = \frac{\rho}{2 \MP^2}.
\eeq
In our case, the largest imaginary magnitude of $\omega$
for the instability is
\begin{equation}
\omega_{\rm inst} = i \frac{\alpha M^3}{4 \MP^2} \equiv i\Gamma.
\end{equation}
Just as the usual Jeans instability is removed by Hubble friction
in an expanding universe, we will see in the next section that for
our case that in a de Sitter background with $H > \Gamma$,
this instability is also removed.

As we will now see, $\Gamma^{-1}$ is also the timescale
over which modifications of gravity take place, at a length scale
of order $m^{-1}$. To see this, we can look at the modification of
the $\Phi_c$ propagator. If we immediately go to the static limit
$\omega \to 0$, we see that the $\Phi_c$ propagator is simply
modified from ${1}/{k^2} \to {1}/{(k^2 - m^2)}$
(remembering that $k$ is the 3-momentum). This looks
like a {\it negative} mass squared for $\Phi_c$ (the sign is dictated
by the requirement of a healthy $\pi$ time kinetic term),
and so gives rise
to an oscillatory sin$(mr)/r$ or cos$(mr)/r$ modification of the
Newtonian potential, rather than the familiar $e^{-mr}/r$ Yukawa
form. However, as we already saw for the spin-dependent force
mediated by $\pi$ exchange, because the $\pi$'s do not move at the
speed of light but instead slow down in their propagation at
larger length scales, retardation effects are very important, and
we cannot immediately go to the static $\omega \to 0$ limit. It
is straightforward to find the $\langle \Phi_c \Phi_c \rangle$
propagator
by inverting the $2 \times 2$ kinetic matrix, yielding
\begin{equation}
\left[1 - \frac{\alpha^2 m^2 k^2}{M^2 \omega^2 - \alpha^2 k^4
+ \alpha^2 m^2 k^2} \right] \times \left(-\frac{1}{k^2} \right).
\end{equation}
The factor in the brackets represents a change in the Newtonian
potential. To get an $O(1)$ modification of gravity, we must have
\begin{equation}
\omega^2 - \frac{\alpha^2}{M^2} k^4 \lsim \alpha^2
\frac{m^2}{M^2} k^2.
\end{equation}
This tells us that we have to be close to ``on-shell" for the
$\pi$ excitations; directly in terms of $\omega$ we then have
\begin{equation}
\left|\omega - \frac{\alpha k^2}{M} \right| \lsim \frac{\alpha
m^2}{M} \sim \Gamma.
\end{equation}
So we see that the width of the region in $\omega$ is small, set
by $\Delta \omega \sim \Gamma$, and therefore we must wait for
a long timescale $t_c \sim \Gamma^{-1}$ to see any
modifications of gravity. It is also easy to see that the natural
length scale of the modification is $r_c \sim m^{-1}$. Indeed, if
we work with dimensionless variables $\hat{\omega}, \hat{k}$
defined by $\omega = 2\Gamma \hat{\omega},\, k = m \hat{k}$,
we see that the modulation factor for the Newtonian potential is
simply
\begin{equation}
\left[1 - \frac{\hat{k}^2}{\hat{\omega}^2 - \hat{k}^4 + \hat{k}^2}
\right]
\end{equation}
and so modifications occur for $\hat{\omega},\hat{k} \sim O(1)$.
We could have also seen that $r_c,t_c$ are the natural length and
time scales for modification also directly from the action. By
further defining $\hat{\Phi} = m \Phi_c$ and $\hat{\pi} =
\Gamma^{-1} \pi_c$, the action can be seen to be $O(1)$ in
all hatted variables.

We therefore conclude that modifications of gravity take place at
a distance $r_c$ and time $t_c$ given by
\begin{equation}
r_c \sim \frac{\MP}{M^2},
\qquad t_c \sim \frac{\MP^2}{M^3 \alpha}.
\end{equation}

The fact that $t_c \gg r_c$ makes sense from a number of points of
view. Note that the scale $r_c$ is completely determined by the
$M^4 h_{00}^2$ part of the lagrangian. However, as we have already
remarked, with only this term, classically we have not modified gravity at
all, only fixed to $h_{00} = 0$ gauge. Only with the other terms
(which lead to the $k^4$ spatial kinetic terms for $\pi$) do we
see that there is a genuine modification of gravity---to see any
change in the potential we then have to wait until both terms
become important, and the timescale is then larger than what one
might expect from the coefficient of the $h_{00}^2$ term alone.

Let us see how the modification works explicitly in an example, by
calculating the effective gravitational potential that is felt by
a test mass, outside a source which turns on at the origin at time
$t=0$, {\it i.e.}, a source
\begin{equation}
\rho_m(r,t) = \delta^3(r) \theta(t).
\end{equation}
Of course energy is really conserved here; this can be thought of
as an approximation to a situation where a cloud of dust quickly
collapses to a point-like mass at the origin. We are interested in
the gravitational force that a test particle feels a distance $r$
away, as a function of time $t>0$. Given the $\Phi_c$ propagator
we have calculated, it is easy to Fourier transform back to
position space to find
\begin{equation}
\Phi(r,t) = -\frac{G}{r} \left[ 1 + I(r,t) \right],
\end{equation}
where $I(r,t)$ is a spatial Fourier integral over the momentum
$k$. Introducing the dimensionless variables
\begin{equation}
u = \frac{k}{m}, \quad R = m r, \quad T =
\frac{\alpha M^3}{2\MP^2} = 2\Gamma t,
\end{equation}
we have
\beq
I(r,t) = \frac{2}{\pi} \Bigg\{ &\int_0^1 du\frac{\sin(u R)}{(u^3 -
u)}
\left(1 - \cosh(T u \sqrt{1 - u^2})\right) \nonumber \\
& + \int_1^\infty du \frac{\sin(u R)}{(u^3 - u)} \left(1 - \cos(T u
\sqrt{u^2 - 1})\right) \Bigg\}.
\eeq

\begin{figure}
 \centering\leavevmode\epsfysize=8cm \epsfbox{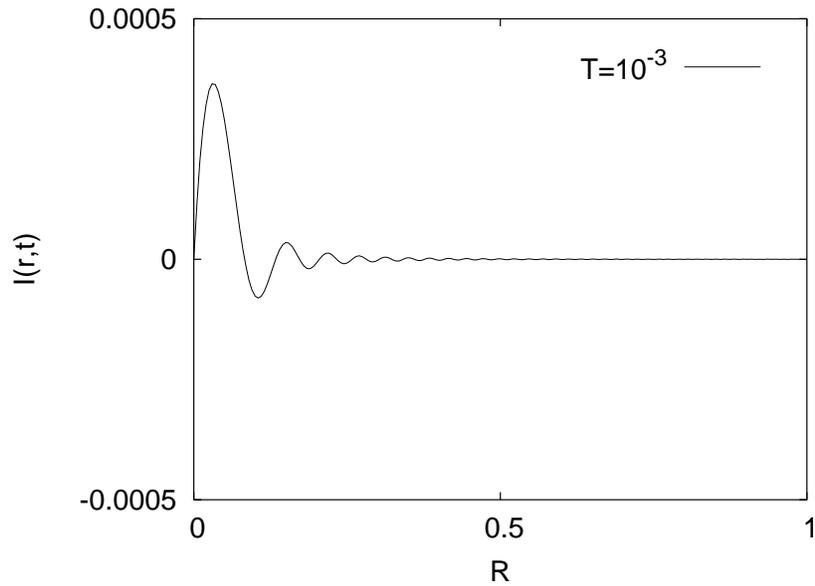}
 \caption{
 The Newtonian potential is $V(r) = -\frac{G}{r} (1 +
 I(r=\frac{R}{m},t=\frac{T}{2\Gamma}))$.
 The function $I(r, t = .001(2\Gamma)^{-1})$ is plotted
 here.
 }
\end{figure}

\begin{figure}
 \centering\leavevmode\epsfysize=8cm \epsfbox{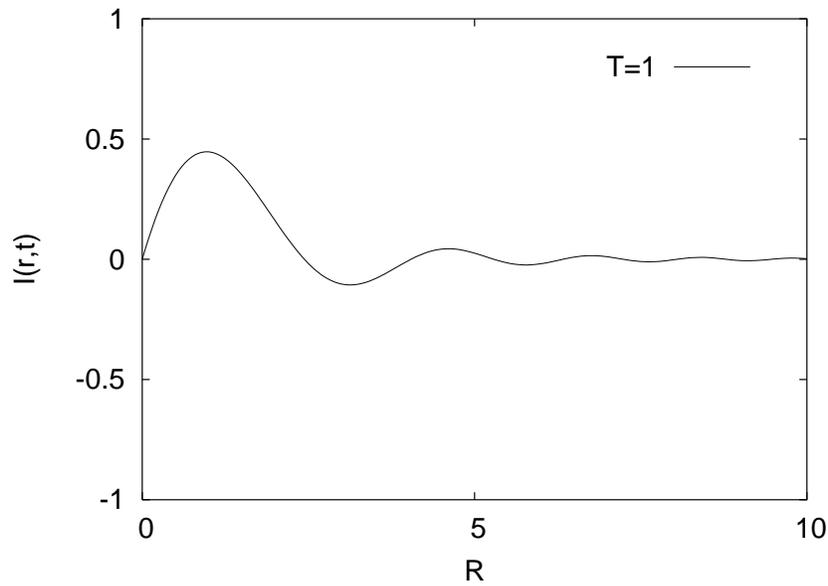}
 \caption{
 Same as the previous figure except for $t = (2\Gamma)^{-1}$.
 Note the axes
 have been rescaled.
 }
\end{figure}

\begin{figure}
 \centering\leavevmode\epsfysize=8cm \epsfbox{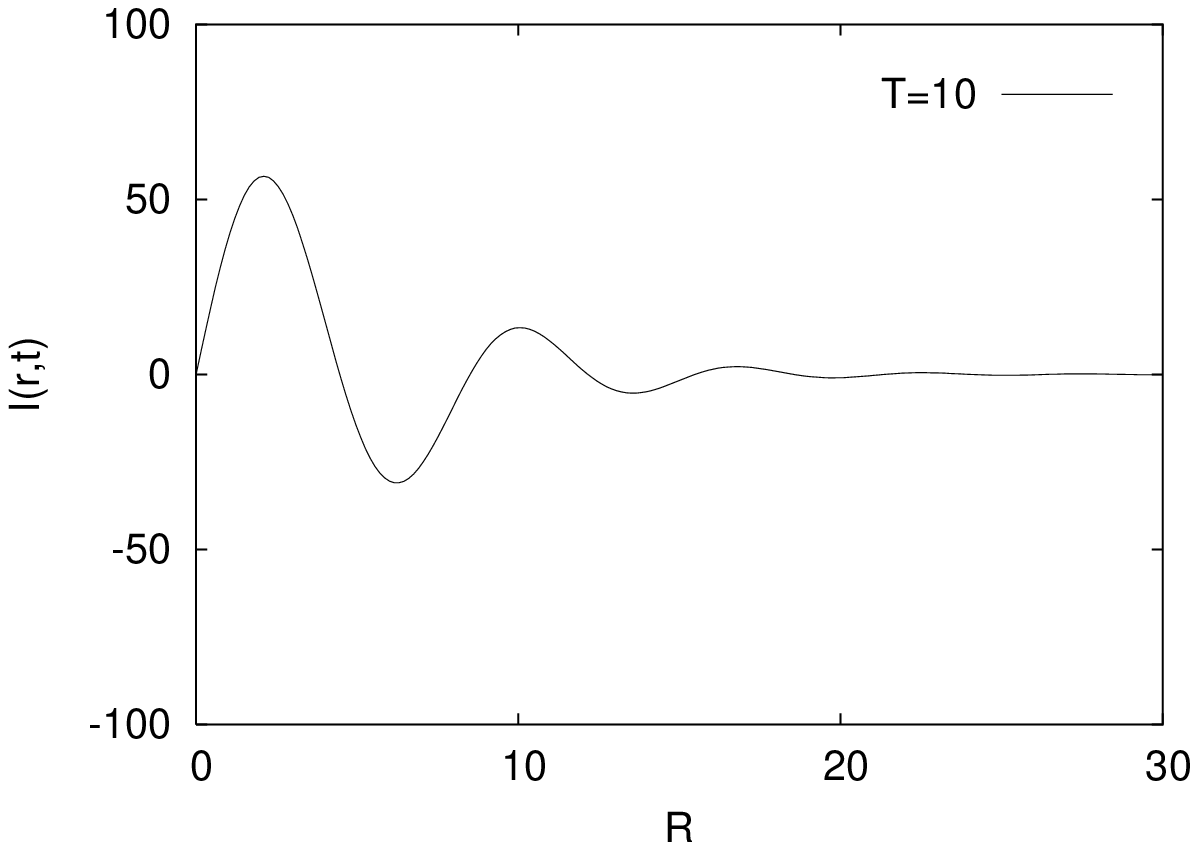}
 \caption{
 Same as the previous figure this time for $t = 5 \Gamma^{-1}$.
 Note again that the axes are rescaled and
 the correction at linear level has become (exponentially)
 larger in amplitude. This is due to the Jeans instability
 of the fluid in flat space}
\end{figure}

For times $ t\ll t_c \, (T \ll 1)$, $I(r,t) \to 0$ and
there is no modification of the standard Newtonian potential.
However, for $t  \gsim t_c \, (T \gsim 1)$, there are modifications.
The presence of the exponentially growing cosh$(T u \sqrt{1 -
u^2})$ term in the first integrand is a reflection of the
Jeans-type instability we found in flat space. For large $T$,
therefore, the first integrand will dominate the integral for
$I(r,t)$. The exponential in the cosh is maximized for $u =
1/\sqrt{2}$; approximating the integrand by a Gaussian around this
point and performing the resulting Gaussian integration then gives
us an excellent approximation for $I(r,t)$:
\begin{equation}
I(r,t) \simeq \frac{2}{\sqrt{{\pi}{T}}} \,
\mbox{exp}\left(-\frac{R^2}{8 T} + \frac{T}{2} \right)
\mbox{sin}\left(\frac{R}{\sqrt{2}}\right).
\end{equation}
So, for $R \ll T$ ($r/r_c \ll t/t_c$), we indeed find an
oscillatory behavior for the Newtonian potential, which is growing
exponentially in time as exp$(T/2)$ as a consequence of the
instability! However, for $R \gg  T$, the correction vanishes; it
takes a time of order $t = t_c \times (r/r_c)$ for any appreciable
changes to happen at distance $r$. The exact behavior of this
Newtonian potential at various time is shown in Fig. 5--7.

Note that for $M \sim 10^{-3}$ eV, which could drive the
acceleration of the universe today, $r_c \sim H_0^{-1}$ is about
the size of the universe,  while $t_c \gg H_0^{-1}$ is much larger
than the age of the universe! Thus, modifications of the Newtonian
potential will occur only at much later time than the present age
of the universe. On the other hand, for $M \sim 10$ MeV, we can
have $t_c \sim H_0^{-1}$ comparable to the present age of the
universe, and these oscillatory modifications of gravity take
place on a distance scale of $\sim 1000$ km!

We close this section by making some brief comments on the
modifications of gravity at non-linear level. In backgrounds with
large gravitational fields such as a neutron star, large
redshift factors will make $\dot{\phi}$ vary by $O(1)$, and we
might wonder whether this leads to large deviations from GR. 
Recall that there is a decoupling limit in which all the
effects of the ghost condensate on the matter sector vanish:
$\MP$ and all matter source energy
densities $\rho_m \to \infty$, while keeping the usual gravitational
scales $\rho_m/\MP^2$ fixed, and also $M$ fixed. 
In this
limit the $\pi$'s have no back-reaction at all on the metric.
Beyond this, it is easy to see that with only the $P(X)$ term in
the lagrangian, if we begin with initial data with
$P^\prime(X=c_*^2)=0$ (as we know we are driven to by Hubble
friction), then $X=c_*^2$ for all time. Thus, in a non-trivial
gravitational potential, $\phi$ adjusts itself so that $X = g^{\mu
\nu} \partial_\mu \phi
\partial_\nu \phi$ is fixed and there is no modification of GR.
Only with the additional terms in the action (that give rise to
the $k^4$ spatial kinetic terms) do we get a deviation from
$X=c_*^2$. As we saw already at linearized level, however, this
effect takes time to build up. At any rate, due to the existence
of the healthy decoupling limit, there is some limit on $M$ which
will be in agreement with all tests of GR. Clearly $M\sim 10^{-3}$
eV will be fine, since in this case even the most naive estimate
of the scale of modification is $M^2/\MP \sim H_0$. A more
detailed investigation is needed to find the precise bounds near
the much larger values of $M \sim 10$ MeV, that lead to deviations
from the linearized Newtonian gravity at timescales of order the
current age of the universe at distances of order $\sim 1000$ km.
It is possible that nonlinear solutions evolve to states outside
the regime of validity of the effective theory, but the same arguments
lead us to expect that this takes very long times.

\section{The Ghost Condensate in de Sitter Space}

In the previous section we studied the ghost condensation
and modification of gravity in a Minkowski background,
and we saw that there is a Jeans-like instability.
As in the case of the Jeans instability for ordinary matter,
we expect that it will
go away in an expanding universe if the expansion rate exceeds
the growth rate of the instability. In this section, we study
the ghost condensation and modification of gravity in
such an expanding de Sitter background.
This is relevant since the current expansion of the universe is
accelerating and we are approaching a de Sitter space.
One could also consider the possibility that ghost condensation is
responsible for inflation in the early universe. In that case,
the fluctuations of ghost condensation in de Sitter space can be
relevant for generating the density perturbations that seed
structure formation in the universe.

It is more convenient to do the analysis using the linearized
equations of motion, so for ease of comparison with our results
from the last section, let us begin by writing the linearized
equations of motion in flat space. In longitudinal gauge, the
Einstein tensor in flat space is
\beq G_{00} &= 2 \nabla^2 \Psi,
\\
G_{0i} &= 2\partial_0 \partial_i \Psi,
\\
G_{ij} &= \delta_{ij} (2\partial_0^2 \Psi + {\nabla^2}(\Phi -
\Psi)) + {\partial_i
\partial_j} (\Psi - \Phi).
\eeq and the equations of motion are
\begin{eqnarray}
G_{00} + \frac{1}{2} M^4 (h_{00} - 2 \partial_0 \pi) = 0, \\
- G_{0i} - \alpha^2 M^2 (\partial_i K) = 0, \\
G_{ij} + \alpha^2 M^2 \partial_0 K \delta_{ij} = 0, \\
-\partial_0 (\partial_0 \pi - \frac{1}{2} h_{00}) -
\frac{\alpha^2 \nabla^2}{M^2} K = 0,
\end{eqnarray}
where we have kept only $h_{00}^2$ and $K^2$ terms and
 set $\MP^2=1/(8\pi G)=1$ for convenience.

{}From the third equation, for $i \neq j$, we find that $\Phi =
\Psi$. Further dropping subleading terms for $M/\MP \ll 1$,
we find
\begin{eqnarray}
\label{einstein_eq00}
2 \nabla^2 \Phi +  M^4 \Phi -  M^4
\partial_0 \pi = 0 ,\\
\label{einstein_eq0i}
-2 \partial_0 \partial_i \Phi - \alpha^2 M^2 \partial_i \nabla^2
\pi = 0 ,\\
\label{einstein_eqij}
2 \partial_0^2 \Phi +  \alpha^2 M^2 \partial_0 \nabla^2
\pi = 0 ,\\
\label{pi_equation}
\partial_0^2 \pi - \partial_0 \Phi + \frac{\alpha^2
\nabla^4}{M^2} \pi = 0.
\end{eqnarray}

The second and third equations in the above both imply that
\begin{equation}
\partial_0 \Phi = -\frac{1}{2} \alpha^2 M^2 \nabla^2 \pi.
\end{equation}
Inserting this into the $\pi$ equation of motion yields
\begin{equation}
\partial_0^2 \pi + \frac{1}{2} \alpha^2 M^2 \nabla^2 \pi +
\alpha^2 \frac{\nabla^4}{M^2} \pi = 0.
\end{equation}
{}From here we can see the modified dispersion relation for $\pi$,
and together with the equation for $\nabla^2 \Phi$, the
modified propagator for $\Phi$ can be derived.

We repeat this exercise in a de Sitter background.
As before, we can use diffeomorphisms
to bring the metric to the following form,
\begin{equation}
ds^2 = (1+2\Phi)dt^2 - a(t)^2 (1-2\Psi)d{\bf x}^2,\quad a(t) = e^{Ht}\, .
\end{equation}
The equations of motion (\ref{einstein_eq00})--(\ref{pi_equation})
are modified to
\begin{eqnarray}
&&-6H \partial_0 \Psi + 2\frac{\nabla^2}{a^2}\Psi-6H^2\Phi+M^4(\Phi-\partial_0 \pi)
= 0 ,\\
&&-\partial_i \left(2H\Phi+2\partial_0 \Psi\right)- \alpha^2 M^2 \partial_i
\left( 3\partial_0 \Psi+\frac{\nabla^2}{a^2}\pi\right) = 0 ,
\\
&&\delta_{ij} \left[ 2H\partial_0 (\Phi +3\Psi) + 2\partial_0^2 \Psi
+ 6H^2(\Phi+\Psi) + \frac{\nabla^2}{a^2}(\Phi-\Psi)\right]
\makebox[0.8in]{}
\nonumber \\
&&-\frac{\partial_i \partial_j}{a^2}(\Phi-\Psi)
 -\delta_{ij}\left[
6H^2 \Psi -\alpha^2 M^2 (\partial_0 +3H)\left(3\partial_0 \Psi
+\frac{\nabla^2}{a^2}\pi \right)\right] = 0 ,
\\
&&(\partial_0 +3H)(\Phi-\partial_0 \pi) - \frac{\alpha^2}{M^2}
\left(3\partial_0 \Psi +\frac{\nabla^2}{a^2}\pi\right) = 0 .
\end{eqnarray}
We consider that the Hubble expansion parameter lies in the
interesting range,
\begin{equation}
\Gamma\left(= \frac{\alpha M^3}{4 \MP^2}\right) \lsim
H \ll m\left(=\frac{M^2}{\sqrt{2} \MP}\right) \ll M .
\end{equation}
For $H \ll \Gamma$ the space-time is effectively flat.
On the other hand, if $H> m$, the length scale of the potential modulation
is outside the horizon and the timescale for $\pi$ to modulate gravity
is way longer than $H^{-1}$, so the interesting modification of gravity
could not have happened.

Again, from the third equation with $i\neq j$, we have $\Phi=\Psi$.
Eliminating $\Phi$ and $\Psi$ from the above equations and neglecting
higher order terms, we obtain
the equation of motion for the $\pi$ field,
\begin{equation}
\label{eq:ds_pi}
\left(\frac{M^4}{2}+ \frac{\nabla^2}{a^2}\right) \partial_0^2 \pi
+ H \left(\frac{M^4}{2}+ 3\frac{\nabla^2}{a^2}\right) \partial_0 \pi
+ \frac{\alpha^2}{M^2}\, \frac{\nabla^2}{a^2}
\left(\frac{M^4}{2}+ \frac{\nabla^2}{a^2}\right)^2 \pi=0.
\end{equation}
By Fourier transforming the spatial dependence, $\nabla^2 \to -k^2$,
the equation of each mode $k$ can be written as
\begin{equation}
\label{eq:ds_pi_k}
\partial_0^2 \pi + H \left(\frac{m^2-3\, \frac{k^2}{a^2}}{m^2-\frac{k^2}{a^2}}
\right)\partial_0 \pi - \frac{4\Gamma^2}{m^4}\, \frac{k^2}{a^2} \left(m^2-
\frac{k^2}{a^2}\right) \pi =0 .
\end{equation}
The $\partial_0 \pi$ term provides the friction (anti-friction) for modes
with $k/a > m$ or $k/a<m/\sqrt{3}$ ($m/\sqrt{3} < k/a < m$). The redshift
will eventually bring each mode to the regime $k/a \ll m/\sqrt{3}$ and
the friction term becomes simply $H \partial_0 \pi$.
For $H\gsim\Gamma$, the friction term dominates and the instability
disappears just like the usual Jeans instability.

To study the modification of gravity due to the ghost condensate,
we should examine the gravitational potential by eliminating $\pi$
instead.
The equation for the gravitational potential $\Phi$ without additional
source is
\begin{equation}
\label{eq:ds_Phi}
\left( \partial_0^2 + 3H \partial_0 +2H^2\right)\Phi
+\frac{\alpha^2}{M^2}\left(\frac{\nabla^2}{a^2}\right)^2 \Phi
+\frac{\alpha^2 M^2}{2}\, \frac{\nabla^2}{a^2} \Phi =0 ,
\end{equation}
or for each momentum mode $k$,
\begin{equation}
\label{eq:ds_Phi_k}
\left( \partial_0^2 + 3H \partial_0 +2H^2\right)\Phi
+\frac{4\Gamma^2}{m^4}\left(\frac{k^4}{a^4}\right) \Phi
-\frac{4\Gamma^2}{m^2}\,\left( \frac{k^2}{a^2}\right) \Phi =0 .
\end{equation}
They are equivalent to Eqs.~(\ref{eq:ds_pi}),
(\ref{eq:ds_pi_k}).
Absence of instability
for $H\gsim\Gamma$ can also be shown with the $\Phi$ equation.
At a large distance ($\gg m^{-1}$) from a local lump of excitations, one
can neglect the last two terms of the equation. The solutions at a
fixed comoving distance $r$ are simply
\begin{equation}
\Phi(r, t) \sim b_1(r)\, e^{-Ht}+ b_2(r)\, e^{-2Ht}.
\end{equation}
They decay at least as fast as the redshifting so there is no growing
potential.

The equation of the gravitational potential with ordinary matter source
can be similarly derived. It is convenient to decompose $\Phi$ into two
parts as
\begin{equation}
 \Phi = \Phi_{\rm GR} + \Phi_{\rm mod},
\end{equation}
where each part satisfies
\begin{equation}
 \frac{\nabla^2}{a^2}\Phi_{\rm GR} = \frac{\rho_{\rm eff}}{2},
  \quad  \rho_{\rm eff} = \rho-4\nabla^2\tilde{p},
\end{equation}
and
\begin{eqnarray}
 \left( \partial_0^2 + 3H \partial_0 +2H^2\right)\Phi_{\rm mod}
  + \frac{4\Gamma^2}{m^4}\left(\frac{\nabla^2}{a^2}\right)^2 \Phi_{\rm mod}
  & + & \frac{4\Gamma^2}{m^2}\, \frac{\nabla^2}{a^2} \Phi_{\rm mod}\nonumber\\
  & = &-\frac{4\Gamma^2}{m^2} \frac{\nabla^2}{a^2}\Phi_{\rm GR},
   \label{eqn:eq-modulation-dS}
\end{eqnarray}
respectively, for the general scalar-type matter source
\begin{equation}
 T_{00} = \rho, \quad
  T_{0i} = \partial_i q, \quad
  a^{-2}T_{ij} = p\delta_{ij} +
  \left(2\nabla_i\nabla_j-\frac{2}{3}\delta_{ij}\nabla^2\right)\tilde{p}.
\end{equation}
To derive the above equations we have used the conservation equations
\begin{eqnarray}
 \dot{\rho} + 3H\rho + 3Hp -\frac{\nabla^2}{a^2}q & = & 0,
  \nonumber\\
 \dot{q} + 3Hq
  - \frac{4}{3}\nabla^2\tilde{p} - p & = & 0.
\end{eqnarray}
Note that $\Phi_{\rm GR}$ is exactly the gravitational potential in general
relativity and, thus, $\Phi_{\rm mod}$ represents the modulation of gravity
due to the ghost condensation. Equation (\ref{eqn:eq-modulation-dS})
shows that the timescale of the modulation is indeed $\Gamma^{-1}$ for
the length scale $m^{-1}$.

To obtain the potential as a function of the physical
distance, we can switch to the physical coordinates $X=a(t)x$, then the
equation for $\Phi_{\rm mod}$ becomes
\begin{eqnarray}
(\partial_0& +& H\, {\bf X}\cdot \nabla_X)^2\Phi_{\rm mod}
+ 3H (\partial_0+H\,{\bf X}\cdot \nabla_X)\Phi_{\rm mod} +2H^2\, \Phi_{\rm mod}
\nonumber\\
&& + \frac{4\Gamma^2}{m^4}\left(\nabla_X^2\right)^2 \Phi_{\rm mod}
+\frac{4\Gamma^2}{m^2}\, {\nabla_X^2} \Phi_{\rm mod}
= -\frac{2\Gamma^2}{m^2}\rho_{\rm eff}
\end{eqnarray}
The modulation of the gravitational potential from a source
also starts at the time
scale $\Gamma^{-1}$ as the $\pi$ field slowly reacts to the gravitational
potential.
If we are interested in the potential at very late time,
we can look for time independent solutions for $\Phi$
by setting $\partial_0=0$ as there is no instability in this case.
Assuming spherical symmetry, the time independent equation is
\begin{eqnarray}
\label{eq:t-indep}
(\sigma^2\partial_\sigma^2+4\sigma\partial_\sigma+2)\Phi_{\rm mod}
+ \frac{4\Gamma^2}{H^2}
\left( \partial_\sigma^4+\frac{4}{\sigma}\partial_\sigma^3\right)\Phi_{\rm mod}
&+&\frac{4\Gamma^2}{H^2}\left(\partial_\sigma^2+
\frac{2}{\sigma}\partial_\sigma\right)\Phi_{\rm mod}\nonumber\\
& = & -\frac{2\Gamma^2}{H^2}\frac{\rho_{\rm eff}}{m^2},
\end{eqnarray}
where $\sigma \equiv m|X|$.

\begin{figure}
 \centering\leavevmode\epsfysize=8cm \epsfbox{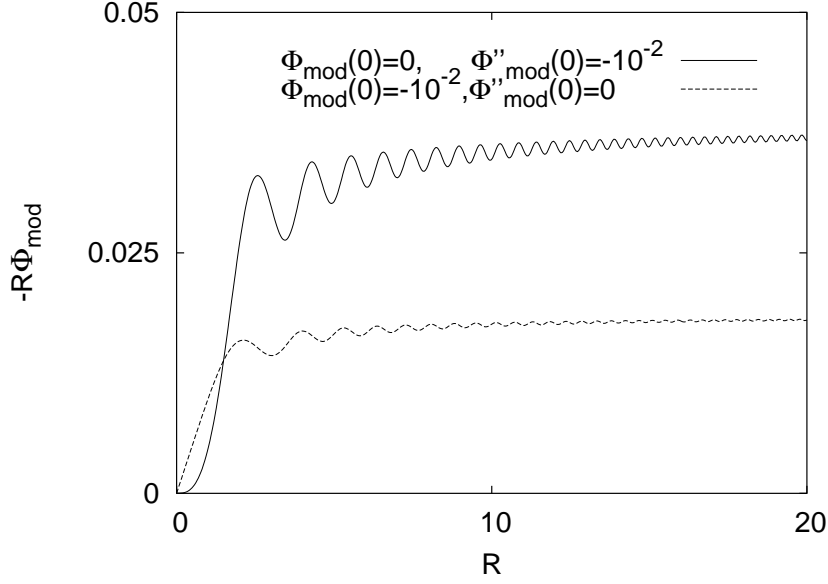}
 \caption{\label{fig:nosource}
 Numerical solutions without matter source
 ($\rho\equiv 0$) for the
 boundary condition ($\Phi_{\rm mod}(0)=0$, $\Phi_{\rm mod}''(0)=-0.01$), and
 ($\Phi_{\rm mod}(0)=-0.01$, $\Phi_{\rm mod}''(0)=0$) with $H/\Gamma=2$.
 }
\end{figure}

We first consider the solutions without matter source and, thus, without
$\Phi_{\rm GR}$. The regularity at the origin $\sigma=0$ requires that all
odd-order derivatives of $\Phi_{\rm mod}$ vanish at $\sigma=0$. Hence, there
are two independent solutions. The two independent solutions are shown
in Fig.~\ref{fig:nosource} for $H/\Gamma=2$, where we plot the potential
multiplied by the distance from the origin.
One can see that apart from the oscillatory modulation, the potential
is proportional to $1/R$ at large $R$. This can be interpreted as
arising from an effective mass produced by a lump of $\pi$ excitations
around the origin. The expansion of the universe keeps these $\pi$
excitations stationary ($\dot{\pi}\neq 0$), resulting in a static
potential. Any linear combination of these two configurations represents
a possible gravitational potential produced by $\pi$ without matter
source, and the effective mass can be either positive or negative which
corresponds to a gravitating or an antigravitating object respectively.

\begin{figure}
 \centering\leavevmode\epsfysize=8cm \epsfbox{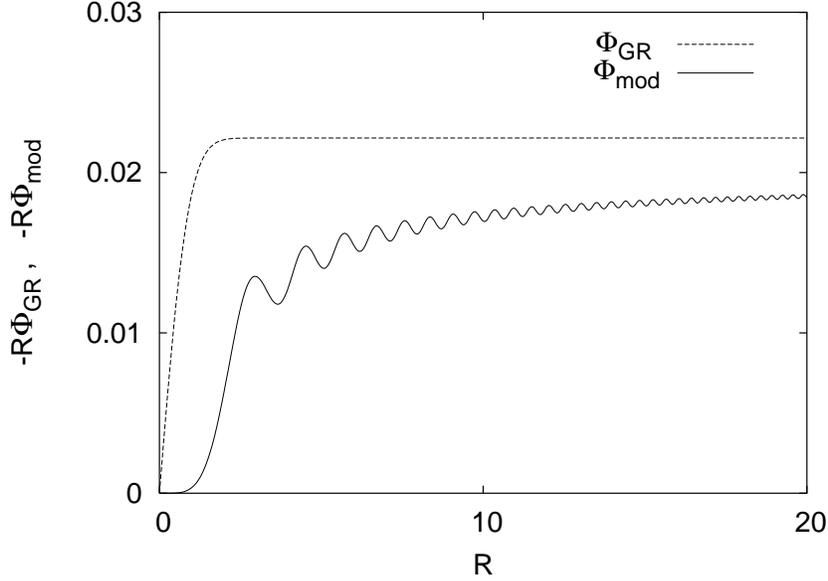}
 \caption{\label{fig:withsource}
 A numerical solution with matter source
 $\rho_{\rm eff}/m^2= 0.1e^{-1.0\sigma^2}$ for the boundary condition
 $\Phi_{\rm mod}(0)=\Phi_{\rm mod}''(0)=0$ and $H/\Gamma=2$.
 }
\end{figure}

A time-independent solution of the potential
with a regular matter source can also be obtained
similarly. A solution with
$\rho_{\rm eff}/m^2=0.1e^{-1.0\sigma^2}$ for $H/\Gamma=2$ is shown
in Fig.~\ref{fig:withsource}.
Again we see the interesting modulation of the $1/R$ potential.
The mass $-R\Phi$~\cite{Misner-Sharp} seen by an observer far away
($m^{-1}\ll R\ll H^{-1}$) is not necessarily the same as the mass of the
matter source due to the screening or anti-screening of the $\pi$
field. A general time-independent solution with the same matter source
is given by the sum of the special solution shown in
Fig.~\ref{fig:withsource} and an arbitrary linear combination of the two
source-free solutions shown in Fig.~\ref{fig:nosource}.
The coefficients of the source-free solutions cannot be determined
by the time independent equation (\ref{eq:t-indep}) but should
be determined by dynamical considerations.


This situation is in stark contrast to the corresponding situation in
general relativity. In general relativity, a spherical symmetric
spacetime outside matter source is locally static and, thus, locally a
Schwarzschild geometry. This fact, known as the Birkhoff
theorem~\cite{Birkhoff}, makes it possible to characterize the geometry
outside the source just by a mass parameter. In general relativity, this
is true even if the geometry inside the matter source is completely
dynamical. Notice also that even in the presence of a usual scalar field,
the Birkhoff theorem holds up to linear perturbation around Minkowski or
de Sitter background, because the scalar field does not couple to gravity
in the linearized level.
(It only appears quadratically or higher order in $T_{\mu\nu}$.)
On the other hand, in ghost
condensation, $\pi$ reacts to the gravitational potential produced by
matter source and slowly modulates the geometry in such a way that the
final stationary configuration depends on the history of the whole
system. This happens even in the linear perturbation level. What
Fig.~\ref{fig:withsource} shows is just one of such possible final
configurations. The reason why this remarkable modulation happens even
in the linear perturbation around a de Sitter background is that we have
a non-vanishing $\dot{\phi}$ in the background.

\section{Discussion and Outlook}

We have presented a consistent modification of gravity in the
infrared. It is useful, though not necessary, to view this arising
from gravity propagating in a ghost condensate. An immediate
application of this model is an alternative to a cosmological
constant for driving a de Sitter phase in the universe. This can be
used to drive the acceleration of the universe today, and can also
be used to for an early inflationary de Sitter phase of the
universe, as will be discussed in \Ref{ghostinfl}. The physics here is
very different from standard slow-roll inflation, and the model
makes sharp predictions about the spectrum of density
perturbations that can be excluded or confirmed in future experiments
\cite{ghostinfl}.

If the standard model fields have direct couplings to the ghost
sector, there are interesting Lorentz-violating signals and
inverse-square law long range forces mediated by excitations of
the ghost condensate. But most interestingly, gravity is modified
in the infrared in a remarkable way by the ghost condensate,
giving rise to the possibility of anti-gravity and
an oscillatory modulation of the
Newtonian potential at late time and large distances.

Our exploration of these models is still in its infancy. The most
pressing question is to determine the current experimental limits
on the parameters of this theory, given the unusual new forces and
modifications of gravity it entails. It would also be interesting
to explore further the possibility raised in section 3 that the
ghost condensate may contribute to both the dark matter and the 
dark energy
of the universe. The physics of non-linear gravitational effects,
and particularly black holes in these backgrounds, is clearly
interesting to explore. 
There are also a number of novel avenues
to explore for early universe cosmology and models with extra
dimensions.
It would also be
interesting to derive our low-energy effective theory from a more
fundamental UV complete theory, either arising from ghost
condensation or in some other way. Finally, can the fact that we
have a model where the ``energy that gravitates" is not the
``particle physics energy" help with the cosmological constant
problem?

\section{Acknowledgments}

We would like to thank Paolo Creminelli, Josh Erlich, Ian Low,
Riccardo Rattazzi, Matthew Schwartz,
and Matias Zaldarriaga for useful discussions.
H.-C.~C. and M.~A.~L. thank Aspen Center for Physics where part of
the work was done. N.~A.-H. and H.-C.~C. are supported by 
NSF grant PHY-0244821, M.~A.~L. is supported by NSF grant PHY-009954 
and S.~M. is supported by NSF grant PHY-0201124. N.~A.-H. is
also supported in part by the David and Lucille Packard foundation. 

\end{document}